\documentclass[pra,byrevtex,superscriptaddress,twocolumn]{revtex4}%
\usepackage{amsfonts}
\usepackage{amsmath}
\usepackage{amssymb}
\usepackage{graphicx}%
\begin{document}
\preprint{ }
\title{Theory of Thermal Motion in Electromagnetically Induced Transparency:\\Diffusion, Doppler, Dicke and Ramsey}
\author{O. Firstenberg}
\affiliation{Department of Physics, Technion-Israel Institute of Technology, Haifa 32000, Israel}
\author{M. Shuker}
\affiliation{Department of Physics, Technion-Israel Institute of Technology, Haifa 32000, Israel}
\author{R. Pugatch}
\affiliation{Department of Physics of Complex Systems, Weizmann Institute of Science,
Rehovot 76100, Israel}
\author{D. R. Fredkin}
\affiliation{Department of Physics, University of California, San Diego, La Jolla,
California 92093}
\author{N. Davidson}
\affiliation{Department of Physics of Complex Systems, Weizmann Institute of Science,
Rehovot 76100, Israel}
\author{A. Ron}
\affiliation{Department of Physics, Technion-Israel Institute of Technology, Haifa 32000, Israel}


\begin{abstract}
We present a theoretical model for electromagnetically induced transparency
(EIT) in vapor, that incorporates atomic motion and velocity-changing
collisions into the dynamics of the density-matrix distribution. Within a
unified formalism we demonstrate various motional effects, known for EIT in
vapor: Doppler-broadening of the absorption spectrum; Dicke-narrowing and
time-of-flight broadening of the transmission window for a finite-sized probe;
Diffusion of atomic coherence during storage of light and diffusion of the
light-matter excitation during slow-light propagation; and Ramsey-narrowing of
the spectrum for a probe and pump beams of finite-size.

\end{abstract}
\maketitle

\section{Introduction}

\begin{figure}
[ptb]
\begin{center}
\includegraphics[
height=7.5806cm,
width=8.1012cm
]%
{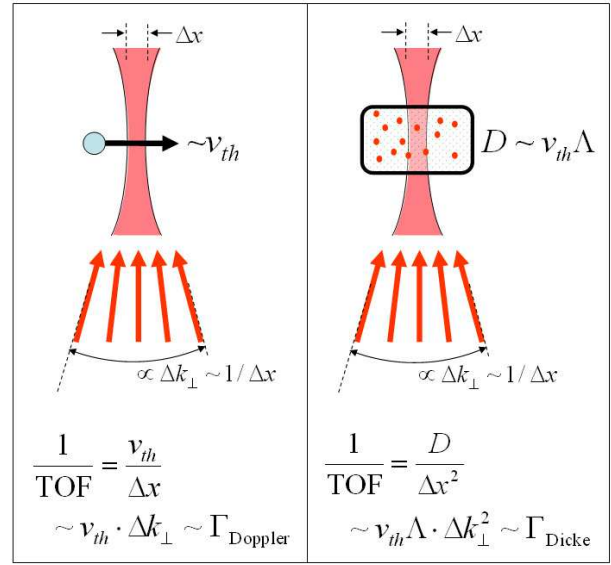}%
\caption{Illustration of time-of-flight (TOF) broadening in the Doppler (left)
and Dicke (right) limits. Assume a beam of width $\Delta x$ in the transverse
plane and width $\Delta k_{\bot}\sim1/\Delta x$ in $k-$space. Atoms with a
transverse velocity $v_{th}$ (left) cross the beam in time $\Delta x/v_{th}$
and cause a TOF broadening of the order of $v_{th}\Delta k_{\bot},$ which is
equal to the well-known Doppler-width. Atoms that undergo diffusion (right),
traverse the beam in average time of $\Delta x^{2}/D$, where $D=v_{th}\Lambda$
is the diffusion coefficient and $\Lambda$ is the mean-free path between
collisions. This results in a TOF broadening of the order of $v_{th}%
\Lambda\Delta k_{\bot}^{2},$ which is the well-known Dicke-width.}%
\label{fig_diffusion_vs_doppler}%
\end{center}
\end{figure}

The Doppler effect, discovered in the mid-19th century, causes a broadening of
spectral lines in thermal media which is linearly proportional to the
radiation wave-vector \cite{Fermi1932}. In 1953 R. H. Dicke predicted that the
Doppler-broadened spectrum can be narrowed due to frequent velocity-changing
collisions \cite{Dicke1953,GalatryPR1961,Nelkin1964}, by a factor proportional
to the ratio between the collisions mean free-path and the radiation
wavelength. This phenomenon, known as \emph{Dicke narrowing}, was observed for
microwave and optical transitions \cite{budker2005,Dutier2003}. When the
motion of the atoms is diffusive, the resulting width is proportional to the
diffusion coefficient and quadratic in the radiation wave-vector
\cite{Nelkin1964,Corey1984}. Therefore it is sometimes referred to as
diffusion-narrowing (of the Doppler profile) or diffusion-broadening (of the
spectrum of a stationary atom). For a finite-size beam, as illustrated in Fig.
\ref{fig_diffusion_vs_doppler}, both the Doppler and the Dicke widths can be
explained as a time-of-flight (TOF) broadening. A comprehensive literature
survey and a theoretical treatment of the Doppler-Dicke problem is presented
by May \cite{May1999}.%

%TCIMACRO{\FRAME{ftbpFU}{8.1012cm}{7.5806cm}{0pt}{\Qcb{Illustration of
%time-of-flight (TOF) broadening in the Doppler (left) and Dicke (right)
%limits. Assume a beam of width $\Delta x$ in the transverse plane and width
%$\Delta k_{\bot}\sim1/\Delta x$ in $k-$space. Atoms with a transverse velocity
%$v_{th}$ (left) cross the beam in time $\Delta x/v_{th}$ and cause a TOF
%broadening of the order of $v_{th}\Delta k_{\bot},$ which is equal to the
%well-known Doppler-width. Atoms that undergo diffusion (right), traverse the
%beam in average time of $\Delta x^{2}/D$, where $D=v_{th}\Lambda$ is the
%diffusion coefficient and $\Lambda$ is the mean-free path between collisions.
%This results in a TOF broadening of the order of $v_{th}\Lambda\Delta k_{\bot
%}^{2},$ which is the well-known Dicke-width.}}{\Qlb{fig_diffusion_vs_doppler}%
%}{diffusion_vs_doppler.eps}{\special{ language "Scientific Word";
%type "GRAPHIC";  maintain-aspect-ratio TRUE;  display "USEDEF";
%valid_file "F";  width 8.1012cm;  height 7.5806cm;  depth 0pt;
%original-width 1.9726in;  original-height 1.8447in;  cropleft "0";
%filename 'diffusion_vs_doppler.eps';file-properties "XNPEU";}}}%
%BeginExpansion
%EndExpansion

Electromagnetically Induced Transparency (EIT) occurs when two light fields, a
$\emph{probe}$ and a \emph{pump}, couple two sublevels of an atomic ground
state manifold to a common upper level. When the Raman resonance conditions
are satisfied, a quantum coherence evolves within the ground state in the
process of coherent population trapping (CPT) \cite{Arimondo96}, inducing a
unique transparency window in the absorption spectrum, accompanied by
dispersion. The spectral width of these features depends primarily on the
ground-state decoherence rates, ranging from the order of several Hz in
cryogenically-cooled crystals \cite{longdell2005} up to the order of several
KHz or MHz in thermal vapor \cite{BollerPRL1991}. A wide variety of phenomena
has been demonstrated utilizing the ultra-narrow EIT resonances, e.g. slow
light \cite{HauNature1999}, stored light \cite{LukinRMP2003}, and non-linear
optics at low light levels \cite{HarrisHauPRL1999}. Having the advantage of
being relatively simple to implement, thermal vapor EIT has been used for
various applications, such as frequency standards \cite{Cyr1993,knappe:1460},
magnetometers \cite{schwindt:6409} and optical delay lines. In this work we
present a comprehensive analytic model for the effect of atomic motion on
vapor EIT, describing broadening and narrowing mechanisms, and the phenomena
of diffusion during slowing and storage of light, within a unified formalism.

As a two-photon process, EIT in room-temperature vapor is susceptible to
residual Doppler broadening, proportional to $\delta q\mathbf{=|q}%
_{1}-\mathbf{q}_{2}|$, where $\mathbf{q}_{1}$ and $\mathbf{q}_{2}$ are the
wave-vectors of the probe and the pump, respectively. However in the presence
of buffer gas, when diffusion dominates the atomic motion, a Dicke-like
narrowing of the Doppler spectrum may occur and the narrowing factor is
proportional to the ratio between the mean-free path and $\delta q^{-1}$
\cite{FirstenbergPRA2007}. In an hyperfine EIT, $\delta q^{-1}$ is of the
order of centimeters and the narrowing factor is of the order $10^{-4},$
completely diminishing the residual Doppler effect and allowing the high
accuracy of EIT-based frequency standards \cite{Cyr1993,WynandsPRA1999}. By
introducing a small angular deviation between the probe and the pump, it is
possible to quantitatively measure the residual-Doppler width and the Dicke
width of Zeeman EIT lines \cite{WeitzPRA2005,ShukerPRA2007}.

For probe and pump beams of \emph{finite size}, the EIT spectrum is subjected
also to TOF broadening \cite{Pfleghaar1993}. Here again, the broadening
reduces to the Dicke-type in the limit of diffusional motion. Nevertheless,
the spectrum is also affected by a more complicated mechanism, recently
denoted as \emph{Ramsey narrowing}, which is attributed to the random-walk of
atoms in and out of the beams \cite{WalsworthJMO2005,WalsworthPRL2006}. The
process in which an atom carrying ground-state coherence leaves the beam,
evolves "in the dark", and then re-enter the beam, is analogous to the Ramsey
method of separated oscillating fields \cite{Ramsey1950}. This process is more
relevant for two-photon phenomena, such as velocity-selective optical pumping,
magneto-optical spectroscopy and CPT/EIT, because of the small frequencies
($c\cdot\delta q$) and long coherence times, that are comparable with the
typical time the atoms spend inside and outside the beam. Ramsey-like features
in such systems were observed in the velocity, time and space domains
\cite{Buhr1986,ZibrovOL2001,ZibrovPRA2001,ClaironPRL2005}. Ramsey narrowing
occurs due to the random distribution of the durations "in the dark",
resulting in a superposition of the spectral Ramsey fringes, that wipes out
the fringes and leaves a single narrow feature in the center. A theoretical
technique of integrating over all possible Ramsey sequences, taking into
account calculated distributions of the durations in and out of the beams,
agrees well with experimental results \cite{WalsworthPRL2006}.

While Ramsey narrowing is an evidence for the diffusion of EIT coherence, a
more explicit demonstration was given in recent storage-of-light experiments,
in which an elaborated spatial profile of the probe field was stored and
retrieved in vapor \cite{PugatchPRL2007,ShukerImaging2007}. The restored
profiles were exactly predicted by assuming that during storage, when no
fields are present, the ground-state coherence undergoes regular diffusion.
Since the coherence is complex, the phase pattern also diffused and
interference effects were observed. Neither a more basic theory that derives
the diffusion equation for the atomic coherence in the absence of fields nor a
prediction for the diffusion-like behavior of slow-light were presented.

The model presented in this paper describes the effect of velocity-changing
collisions on the position and velocity dependence of the atomic density
matrix. Assuming a Boltzmann-like relaxation in velocity space, we write in
section \ref{ch_EOM} the dynamics equations for the internal and the external
atomic motion, under the weak-probe approximation. We then derive the
equations of motion for the slowly-varying envelopes of the atomic coherences
and the probe's field, while the pump's envelope is assumed stationary. In
section \ref{ch_general_solution} we formally solve these equations and derive
the probe's susceptibility, for the \emph{general} case and for a plane-wave
pump. By \emph{general} we mean that neither the "Doppler" nor the "Dicke"
limits are taken \cite{May1999} for the one-photon or the two-photon
transitions. This model extends our previous study on Dicke-narrowing
\cite{FirstenbergPRA2007} in the following aspects: it avoids the use of the
low-contrast approximation, incorporating the power-broadening effect; it
allows for a non-planar (finite-size) probe beam, introducing a
\emph{wave-vector filter}; it is time-dependent and thus allows for the
propagation of a probe package. We note also that the assumption of Gaussian
distribution of atomic trajectories is avoided here (see Eq. (9) in Ref.
\cite{FirstenbergPRA2007}). In section \ref{ch_dicke_diff} we consider the
realistic regime, where the velocity relaxation-rate is large enough to cause
Dicke-narrowing of the two-photon transition. In this regime we derive a
diffusion equation for the density-matrix distribution during storage-of-light
and afterwards calculate the dynamics in the presence of the fields. For the
case of plane-wave pump we recover the Dicke-Diffusion absorption spectrum and
arrive at a diffusion-diffraction equation for the slowly propagating envelope
of the probe. For the case of finite-size pump and probe, analyzed in section
\ref{ch_ramsey}, we solve the diffusion equations and retrieve the
Ramsey-narrowed absorption spectrum. Note that the latter is done in a
steady-state approach, rather then by averaging over all possible atomic trajectories.

\section{Equations of Motion\label{ch_EOM}}

\subsection{Atom-Field Interaction}

\begin{figure}
[ptb]
\begin{center}
\includegraphics[
trim=0.000000in 0.000000in 0.000762in 0.000000in,
height=2.5313in,
width=3.0727in
]%
{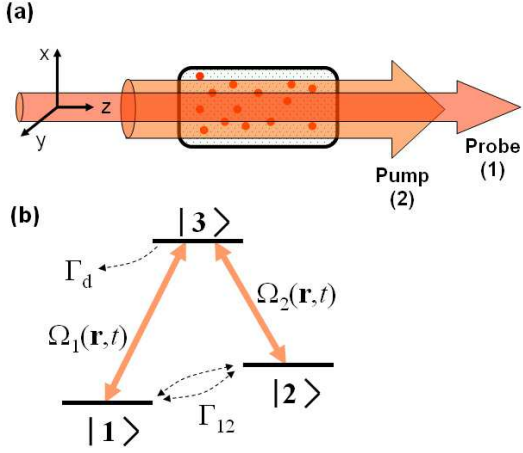}%
\caption{(a) A probe beam and a pump beam, with a finite envelope in space,
propagate through the vapor cell. The $z-$axis is chosen perpendicular to the
probe's direction and the $x$ and $y$ axes form the \emph{transverse plane}.
(b) Atomic levels diagram. $\Omega_{1}$ and $\Omega_{2}$ are the Rabi
frequencies of the probe and the pump, respectively. $\Gamma_{d}$ and
$\Gamma_{12}$ are the decoherence rates of the optical and the ground-state
transitions.}%
\label{fig_appartus_and_levels}%
\end{center}
\end{figure}

We consider three-level atoms in a $\Lambda-$configuration, with an upper
state $\left\vert 3\right\rangle $ and two lower states, $\left\vert
2\right\rangle $ and $\left\vert 1\right\rangle ,$ as depicted in Fig.
\ref{fig_appartus_and_levels}. The atoms interact with two external, classical
electromagnetic fields, propagating in time $t$ and space $\mathbf{r}$,
\begin{equation}
\mathbf{E}_{s}\left(  \mathbf{r},t\right)  =\Re e\left\{  \frac{\hbar}%
{\mu_{3s}}\mathbf{\epsilon}_{s}\Omega_{s}\left(  \mathbf{r},t\right)
\right\}  ~~\text{for }s=1,2,\label{R1-2}%
\end{equation}
with%
\begin{equation}
\Omega_{s}\left(  \mathbf{r},t\right)  =\tilde{\Omega}_{s}\left(
\mathbf{r},t\right)  e^{-i\omega_{s}t}e^{i\mathbf{q}_{s}\cdot\mathbf{r}%
},\label{Rl-11}%
\end{equation}
where $\mathbf{\epsilon}_{s}$\ and $\mathbf{q}_{s}$ are the polarization
vector and wave vector of the probe ($s=1$) and the pump ($s=2$); $\omega
_{s}=c\left\vert \mathbf{q}_{s}\right\vert $; $\mu_{3s}=\left\langle
3\right\vert \mathbf{d\cdot\epsilon}_{s}\left\vert s\right\rangle $\ is the
$3\rightarrow s$\ transition dipole moment matrix element; and $\widetilde
{\Omega}_{s}\left(  \mathbf{r},t\right)  $\ is the \textit{slowly varying
envelope in time} of the Rabi frequency, satisfying $|\frac{\partial}{\partial
t}\tilde{\Omega}_{s}\left(  \mathbf{r},t\right)  |\ll\omega_{s}|\tilde{\Omega
}_{s}\left(  \mathbf{r},t\right)  |$.%

%TCIMACRO{\FRAME{ftbpFU}{3.0727in}{2.5313in}{0pt}{\Qcb{(a) A probe beam and a
%pump beam, with a finite envelope in space, propagate through the vapor cell.
%The $z-$axis is chosen perpendicular to the probe's direction and the $x$ and
%$y$ axes form the \emph{transverse plane}. (b) Atomic levels diagram.
%$\Omega_{1}$ and $\Omega_{2}$ are the Rabi frequencies of the probe and the
%pump, respectively. $\Gamma_{d}$ and $\Gamma_{12}$ are the decoherence rates
%of the optical and the ground-state transitions.}}%
%{\Qlb{fig_appartus_and_levels}}{apparatus_and_levels.eps}%
%width 3.0727in;  height 2.5313in;  depth 0pt;  original-width 1.9043in;
%original-height 1.5653in;  cropleft "0";  croptop "1";  cropright "0.9996";
%cropbottom "0";  filename 'apparatus_and_levels.eps';file-properties "XNPEU";}%
%}}%
%BeginExpansion
%EndExpansion

First we consider a \textit{single atomic system}, the $i-$th\ atom, one of
many\textit{\ identical} particles. The center of mass position $\mathbf{r}%
_{i}\left(  t\right)  $\ enters the internal dynamics of the atom because the
external fields must be evaluated there. We denote the $ss^{\prime}$ reduced
density-matrix element of the $i-$th atom as $\rho_{ss^{\prime}}^{i}\left(
t\right)  .$ For a \emph{weak probe}\textit{,} namely when the probe field is
weak enough to cause $\rho_{31}^{i}\ll\rho_{11}^{i},$ the equations of motion
of $\rho_{31}^{i}$ and $\rho_{21}^{i}$ are (cf. \cite{FirstenbergPRA2007})
\begin{align}
\frac{\partial}{\partial t}\rho_{31}^{i}  & =-\Gamma_{d}\rho_{31}^{i}%
-i\omega_{31}\rho_{31}^{i}+i\Omega_{2}\left(  \mathbf{r}_{i},t\right)
\rho_{21}^{i}\nonumber\\
& +i\Omega_{1}\left(  \mathbf{r}_{i},t\right)  \rho_{11}^{i,\text{(eq)}%
},\label{ul-rl-1}\\
\frac{\partial}{\partial t}\rho_{21}^{i}  & =-\Gamma_{21}\rho_{21}^{i}%
-i\omega_{21}\rho_{21}^{i}+i\Omega_{2}^{\ast}\left(  \mathbf{r}_{i},t\right)
\rho_{31}^{i}.\nonumber
\end{align}
Here $\omega_{31}$ and $\Gamma_{d}$ are the frequency and the decoherence rate
of the $3\rightarrow1$ transition; $\omega_{21}$ and $\Gamma_{21}$ are the
frequency and the decoherence rate of the $2\rightarrow1$ transition;
$\rho_{31}^{i}$ is linear in the probe field; and the equilibrium density
matrix in the absence of the probe $\left(  \Omega_{1}=0\right)  $ is
\begin{equation}
\rho_{ss^{\prime}}^{i,\text{(eq)}}=%
\begin{cases}
1 & ~s=s^{\prime}=1\\
0 & ~\text{otherwise}.
\end{cases}
\label{rll-1}%
\end{equation}

We consider the case of \emph{non-depleted pump}, i.e. we assume that the
pump's envelope is constant in time and is given by $\tilde{\Omega}_{2}\left(
\mathbf{r},t\right)  =\tilde{\Omega}_{2}\left(  \mathbf{r}\right)  .$ The wave
equation of the probe field is%
\begin{equation}
\left(  \nabla^{2}-\frac{1}{c^{2}}\frac{\partial^{2}}{\partial t^{2}}\right)
\mathbf{E}_{1}\left(  \mathbf{r},t\right)  =\frac{4\pi}{c^{2}}\frac
{\partial^{2}}{\partial t^{2}}\mathbf{P}_{31}\left(  \mathbf{r},t\right)
,\label{Ert-1}%
\end{equation}
where $\mathbf{P}_{31}\left(  \mathbf{r},t\right)  =\Re e\left\{
\mathbf{\tilde{P}}_{31}\left(  \mathbf{r},t\right)  e^{-i\omega_{1}%
t}e^{i\mathbf{q}_{1}\cdot\mathbf{r}}\right\}  $\ is the contribution of the
$3\rightarrow1$ transition to the \textit{expectation value} of the
polarization density, and $\mathbf{\tilde{P}}_{31}\left(  \mathbf{r},t\right)
$ is the slowly varying polarization. With Eq.(\ref{R1-2}) we have%
\begin{equation}
\left(  \frac{\partial}{\partial t}+c\frac{\partial}{\partial z}-i\frac
{c}{2q_{1}}\nabla^{2}\right)  \tilde{\Omega}_{1}\left(  \mathbf{r},t\right)
=i\frac{g}{\mu_{31}^{\ast}}\mathbf{\epsilon}_{1}\cdot\mathbf{\tilde{P}}%
_{31}\left(  \mathbf{r},t\right)  ,\label{Ert-3}%
\end{equation}
where $g=2\pi\omega_{1}\left\vert \mu_{31}\right\vert ^{2}/\hbar$ is a
coupling constant and $\nabla^{2}$ is the three-dimensional Laplacian
operator. To obtain Eq.(\ref{Ert-3}) we neglected the second order temporal
derivatives of the envelopes and chose, without loss of generality,
$\mathbf{q}_{1}=\mathbf{\hat{z}}q_{1}$, where $\mathbf{\hat{z}}$ is a unit
vector in the $z-$direction (see Fig. \ref{fig_appartus_and_levels}.a). The
second order spatial derivatives are retained to allow the description of
finite-size beams.

\subsection{Boltzmann-like Equations}

We consider a dilute gas of active atoms in the presence of a noble buffer-gas
that causes frequent velocity-changing coherence-preserving collisions. We
introduce a generalized density-matrix distribution function in space and
velocity%
\begin{equation}
\rho_{ss^{\prime}}=\rho_{ss^{\prime}}(\mathbf{r},\mathbf{v},t)=\sum_{i}%
\rho_{ss^{\prime}}^{i}\left(  t\right)  \delta\left(  \mathbf{r}%
-\mathbf{r}_{i}\left(  t\right)  \right)  \delta\left(  \mathbf{v}%
-\mathbf{v}_{i}\left(  t\right)  \right)  ,\label{rs-1}%
\end{equation}
where the time dependence of $\rho_{ss^{\prime}}^{i}\left(  t\right)  $\ is
due to the internal dynamics only, considered earlier in the single-particle
density-matrix. We express%
\begin{align*}
\frac{\partial}{\partial t}\rho_{ss^{\prime}}  & =\sum_{i}\left(
\frac{\partial}{\partial t}\rho_{ss^{\prime}}^{i}\right)  \delta\left(
\mathbf{r}-\mathbf{r}_{i}\left(  t\right)  \right)  \delta\left(
\mathbf{v}-\mathbf{v}_{i}\left(  t\right)  \right) \\
& +\sum_{i}\rho_{ss^{\prime}}^{i}\frac{d\mathbf{r}_{i}}{dt}\mathbf{\cdot
}\left[  \frac{\partial}{\partial\mathbf{r}_{i}}\delta\left(  \mathbf{r}%
-\mathbf{r}_{i}\left(  t\right)  \right)  \right]  \delta\left(
\mathbf{v}-\mathbf{v}_{i}\left(  t\right)  \right) \\
& +\sum_{i}\rho_{ss^{\prime}}^{i}\delta\left(  \mathbf{r}-\mathbf{r}%
_{i}\left(  t\right)  \right)  \frac{d\mathbf{v}_{i}}{dt}\mathbf{\cdot}\left[
\frac{\partial}{\partial\mathbf{v}_{i}}\delta\left(  \mathbf{v}-\mathbf{v}%
_{i}\left(  t\right)  \right)  \right]
\end{align*}
or%
\begin{align}
& \frac{\partial}{\partial t}\rho_{ss^{\prime}}+\mathbf{v\cdot}\frac{\partial
}{\partial\mathbf{r}}\rho_{ss^{\prime}}+\left[  \frac{\partial}{\partial
t}\rho_{ss^{\prime}}\right]  _{\operatorname{col}.}\nonumber\\
& =\sum_{i}\left(  \frac{\partial}{\partial t}\rho_{ss^{\prime}}^{i}\right)
\delta\left(  \mathbf{r}-\mathbf{r}_{i}\left(  t\right)  \right)
\delta\left(  \mathbf{v}-\mathbf{v}_{i}\left(  t\right)  \right)
,\label{rs-3}%
\end{align}
where $\left[  \frac{\partial}{\partial t}\rho_{ss^{\prime}}\right]
_{\operatorname{col}.}$ is the effect of collisions. Notice that the
density-matrix distribution function is classical, as far as its external
motion is concerned, and quantum mechanical in its internal atomic motion. The
function $\rho_{ss^{\prime}}(\mathbf{r},\mathbf{v},t)d^{3}rd^{3}v$ is
interpreted as the probability to find an atom of density-matrix element
$\rho_{ss^{\prime}}$ at the time $t,$\ near position $\mathbf{r,}$\ within a
volume element $d^{3}r,$ with velocity $\mathbf{v}$\textbf{,}\ within a
velocity volume element $d^{3}v.$

With this interpretation we understand the effect of collisions as relaxation
to thermal equilibrium of the center of mass, and we shall take it into
account using the Boltzmann relaxation method. Introducing the density%
\begin{equation}
R_{ss^{\prime}}(\mathbf{r},t)=\int d^{3}v\rho_{ss^{\prime}}(\mathbf{r}%
,\mathbf{v},t),\label{rs-4}%
\end{equation}
which is interpreted as the number of atoms with $\rho_{ss^{\prime}}$ per unit
volume near $\mathbf{r}$ in space, we assume that the distribution
$\rho_{ss^{\prime}}(\mathbf{r},\mathbf{v},t)$ relaxes to equilibrium as
\cite{Nelkin1964}%
\begin{equation}
\left[  \frac{\partial}{\partial t}\rho_{ss^{\prime}}(\mathbf{r}%
,\mathbf{v},t)\right]  _{\operatorname{col}.}=-\gamma\left[  \rho_{ss^{\prime
}}(\mathbf{r},\mathbf{v},t)-R_{ss^{\prime}}(\mathbf{r},t)F\left(
\mathbf{v}\right)  \right]  .\label{rs-5}%
\end{equation}
Here $\gamma$ is the relaxation rate in time, proportional to the collisions
rate, and $F\left(  \mathbf{v}\right)  $ is the thermal equilibrium Boltzmann
distribution function in velocity space,
\begin{equation}
F\left(  \mathbf{v}\right)  =\frac{1}{\left(  2\pi v_{\text{th}}^{2}\right)
^{3/2}}e^{-v^{2}/(2v_{\text{th}}^{2})}~\ ~~~;~~~~v_{\text{th}}^{2}=\frac
{k_{B}T}{m},\label{Fv}%
\end{equation}
where $k_{B}$ is the Boltzmann constant, $T$ is the temperature, $m$ is the
atomic mass and $v_{\text{th}}$ is the thermal velocity. Here the internal
motion, and the external one, are completely separated. The collisions with
the buffer gas affect only the external degrees of freedom and are assumed
much frequent than the collisions with the active atoms. If the buffer gas
particles affect the internal motion, e.g. by pressure-broadening
\cite{Corey1984}, it is taken into account in the atomic decay rates. The
collision term, Eq.(\ref{rs-5}), neglects partial wave scattering for $l>0$
and also neglects the energy dependence of the scattering cross-section. From
Eqs. (\ref{ul-rl-1}-\ref{rll-1},\ref{rs-3}) and the Boltzmann collision term
of Eq.(\ref{rs-5}), we observe that%
\begin{equation}
\sum_{i}\rho_{11}^{i,\text{(eq)}}\delta\left(  \mathbf{r}-\mathbf{r}%
_{i}\left(  t\right)  \right)  \delta\left(  \mathbf{v}-\mathbf{v}_{i}\left(
t\right)  \right)  =n_{0}F\left(  \mathbf{v}\right)  ,
\end{equation}
where $n_{0}$\ is the gas particle density, and the equations of motion are:%
\begin{align}
\left(  \frac{\partial}{\partial t}+\Gamma_{d}+i\omega_{31}+\mathbf{v\cdot
}\frac{\partial}{\partial\mathbf{r}}\right)  \rho_{31}-i\Omega_{2}%
(\mathbf{r})\rho_{21}  & \label{ul-rl-3}\\
-i\Omega_{1}(\mathbf{r},t)n_{0}F\left(  \mathbf{v}\right)  +\gamma\left[
\rho_{31}-R_{31}(\mathbf{r},t)F\left(  \mathbf{v}\right)  \right]   &
=0,\nonumber\\
\left(  \frac{\partial}{\partial t}+\Gamma_{21}+i\omega_{21}+\mathbf{v\cdot
}\frac{\partial}{\partial\mathbf{r}}\right)  \rho_{21}-i\Omega_{2}^{\ast
}(\mathbf{r})\rho_{31}  & \nonumber\\
+\gamma\left[  \rho_{21}-R_{21}(\mathbf{r},t)F\left(  \mathbf{v}\right)
\right]   & =0,\nonumber
\end{align}
with $\rho_{21}=\rho_{21}(\mathbf{r},\mathbf{v},t)$ and $\rho_{31}=\rho
_{31}(\mathbf{r},\mathbf{v},t)$. We note that the above \emph{semiclassical}
analysis of the dynamics has a quantum-mechanical equivalent, which will be of
significance in a regime where the atomic recoil velocity is comparable with
the thermal velocity.

\subsection{Envelope Equations}

Since the probe field propagates through the cell with a slowly varying
envelope, $\rho_{31}(\mathbf{r},\mathbf{v},t)$\ and $\rho_{21}(\mathbf{r}%
,\mathbf{v},t)$\ can be expressed as%
\begin{align}
\rho_{31}(\mathbf{r},\mathbf{v},t)  & =\tilde{\rho}_{31}(\mathbf{r}%
,\mathbf{v},t)e^{-i\omega_{1}t}e^{i\mathbf{q}_{1}\cdot\mathbf{r}}%
,\label{rs-6}\\
\rho_{21}(\mathbf{r},\mathbf{v},t)  & =\tilde{\rho}_{21}(\mathbf{r}%
,\mathbf{v},t)e^{-i\left(  \omega_{1}-\omega_{2}\right)  t}e^{i\left(
\mathbf{q}_{1}-\mathbf{q}_{2}\right)  \cdot\mathbf{r}},\nonumber
\end{align}
where $\tilde{\rho}_{31}$ and $\tilde{\rho}_{21}$ are slowly varying in space
and time. Similarly we introduce the slowly varying densities, $\tilde{R}%
_{21}(\mathbf{r},t)$ and $\tilde{R}_{31}(\mathbf{r},t),$ and express the
expectation value of the polarization density, $\mathbf{\tilde{P}}_{31}\left(
\mathbf{r},t\right)  $, in terms of the density $\tilde{R}_{31}\left(
\mathbf{r},t\right)  $ as
\begin{equation}
\mathbf{\epsilon}_{1}\cdot\mathbf{\tilde{P}}_{31}\left(  \mathbf{r},t\right)
=\mu_{31}^{\ast}\tilde{R}_{31}\left(  \mathbf{r},t\right)  .
\end{equation}
With the \emph{one-photon detuning, }$\Delta_{1}=\omega_{1}-\omega_{31},$ and
the \emph{two-photon Raman detuning, }$\Delta=\omega_{1}-\omega_{2}%
-\omega_{21},$ we define
\begin{subequations}
\label{dl-dr}%
\begin{align}
\xi_{1}  & =\Delta_{1}-\mathbf{q}_{1}\mathbf{\cdot v}+i\left(  \Gamma
_{d}+\gamma\right)  ,\label{dl-3}\\
\xi_{2}  & =\Delta-\left(  \mathbf{q}_{1}-\mathbf{q}_{2}\right)  \mathbf{\cdot
v}+i\left(  \Gamma_{21}+\gamma\right)  .\label{dr1}%
\end{align}
and write Eqs.(\ref{ul-rl-3}) and Eq.(\ref{Ert-3}) as
\end{subequations}
\begin{subequations}
\label{ul-rl-4}%
\begin{align}
\left(  \frac{\partial}{\partial t}+\mathbf{v\cdot}\frac{\partial}%
{\partial\mathbf{r}}-i\xi_{1}\right)  \tilde{\rho}_{31}(\mathbf{r}%
,\mathbf{v},t)-\gamma\tilde{R}_{31}(\mathbf{r},t)F\left(  \mathbf{v}\right)
& \label{ul-4}\\
-i\tilde{\Omega}_{2}(\mathbf{r})\tilde{\rho}_{21}(\mathbf{r},\mathbf{v}%
,t)-i\tilde{\Omega}_{1}(\mathbf{r},t)n_{0}F\left(  \mathbf{v}\right)   &
=0,\nonumber\\
\left(  \frac{\partial}{\partial t}+\mathbf{v\cdot}\frac{\partial}%
{\partial\mathbf{r}}-i\xi_{2}\right)  \tilde{\rho}_{21}(\mathbf{r}%
,\mathbf{v},t)-\gamma\tilde{R}_{21}(\mathbf{r},t)F\left(  \mathbf{v}\right)
& \label{rl-4}\\
-i\tilde{\Omega}_{2}^{\ast}(\mathbf{r})\tilde{\rho}_{31}(\mathbf{r}%
,\mathbf{v},t)  & =0,\nonumber
\end{align}
and
\end{subequations}
\begin{equation}
\left(  \frac{\partial}{\partial t}+c\frac{\partial}{\partial z}-i\frac
{c}{2q_{1}}\nabla^{2}\right)  \tilde{\Omega}_{1}\left(  \mathbf{r},t\right)
=ig\tilde{R}_{31}\left(  \mathbf{r},t\right)  .\label{Ert-4}%
\end{equation}
Eqs.(\ref{ul-rl-4}) and (\ref{Ert-4}) compose the full set of equations of
motion for the slowly varying envelopes.

Finally, in sections \ref{ch_general_solution} and \ref{ch_dicke_diff} we
study the case of a stationary, plane-wave pump, $\tilde{\Omega}_{2}\left(
\mathbf{r}\right)  =\Omega_{2}$. For this case it is convenient introduce the
Fourier transformation\ and replace the slowly varying time dependent and
$\mathbf{r}-$dependent functions by%
\begin{equation}
f\left(  \mathbf{r},t\right)  =\int_{-\infty}^{\infty}\frac{d^{3}k}{2\pi
}e^{i\mathbf{kr}}\int_{-\infty}^{\infty}\frac{d\omega}{2\pi}e^{-i\omega
t}f\left(  \mathbf{k},\omega\right) \label{ft-1}%
\end{equation}
and write Eqs.(\ref{ul-rl-4}) as
\begin{subequations}
\label{ul-rl-5}%
\begin{align}
\left(  \omega-\mathbf{k\cdot v}+\xi_{1}\right)  \tilde{\rho}_{31}%
(\mathbf{k},\mathbf{v},\omega)-i\gamma\tilde{R}_{31}\left(  \mathbf{k}%
,\omega\right)  F\left(  \mathbf{v}\right)   & \label{ul-5}\\
+\Omega_{2}\tilde{\rho}_{21}(\mathbf{k},\mathbf{v},\omega)+\tilde{\Omega}%
_{1}(\mathbf{k},\omega)n_{0}F\left(  \mathbf{v}\right)   & =0,\nonumber\\
\left(  \omega-\mathbf{k\cdot v}+\xi_{2}\right)  \tilde{\rho}_{21}%
(\mathbf{k},\mathbf{v},\omega)-i\gamma\tilde{R}_{21}(\mathbf{k},\omega
)F\left(  \mathbf{v}\right)   & \label{rl-5}\\
+\Omega_{2}^{\ast}\tilde{\rho}_{31}(\mathbf{k},\mathbf{v},\omega)  &
=0,\nonumber
\end{align}
and Eq.(\ref{Ert-4}) as
\end{subequations}
\begin{equation}
\left(  ik_{z}-i\frac{\omega}{c}+i\frac{k^{2}}{2q_{1}}\right)  \tilde{\Omega
}_{1}\left(  \mathbf{k},\omega\right)  =i\frac{g}{c}\tilde{R}_{31}%
(\mathbf{k},\omega).\label{Ert-5}%
\end{equation}
\ 

\section{General Solution\label{ch_general_solution}}

We consider the EIT medium in a cell, with a probe beam of finite width in the
transverse plane $\left(  x,y\right)  $, propagating along the $z-$axis in the
direction of $\mathbf{q}_{1}=\mathbf{\hat{z}}q_{1}$. The variation of
$\tilde{\Omega}_{1}$ in space is much slower than $2\pi/q_{1}$ and the
variation in time is much slower than $2\pi/\omega_{1}$. The pump is a
plane-wave, $\Omega_{2}$, propagating with a wave-vector $\mathbf{q}_{2}$.
Here we present the general solution for the probe field propagation inside
the cell, i.e. away from its boundaries, for any relaxation rate $\gamma$,
without taking the "Doppler" or the "Dicke" limits.

We start from Eqs.(\ref{ul-rl-5}) and formally solve for $\tilde{\rho}_{31}$
and $\tilde{\rho}_{21}$:%
\begin{equation}
\tilde{\rho}_{31}=\frac{F\left(  \mathbf{v}\right)  }{\xi_{d}}\left\vert
\begin{array}
[c]{cc}%
i\gamma\tilde{R}_{31}-\tilde{\Omega}_{1}(\mathbf{k},\omega)n_{0} & \Omega
_{2}\\
i\gamma\tilde{R}_{21} & \omega-\mathbf{kv}+\xi_{2}%
\end{array}
\right\vert ,
\end{equation}
and%
\begin{equation}
\tilde{\rho}_{21}=\frac{F\left(  \mathbf{v}\right)  }{\xi_{d}}\left\vert
\begin{array}
[c]{cc}%
\omega-\mathbf{k\cdot v}+\xi_{1} & i\gamma\tilde{R}_{31}-\tilde{\Omega}%
_{1}(\mathbf{k},\omega)n_{0}\\
\Omega_{2}^{\ast} & i\gamma\tilde{R}_{21}%
\end{array}
\right\vert ,\label{ur-7}%
\end{equation}
where $|\cdots|$ stands for matrix-determinant and%
\begin{equation}
\xi_{d}=\left(  \omega-\mathbf{k\cdot v}+\xi_{1}\right)  \left(
\omega-\mathbf{k\cdot v}+\xi_{2}\right)  -|\Omega_{2}|^{2}.\label{D_d-1}%
\end{equation}
Integrating over velocity, we get coupled equations for $\tilde{R}_{31}$ and
$\tilde{R}_{21},$
\begin{equation}
\left[
\begin{array}
[c]{cc}%
1-i\gamma G_{1} & i\gamma\tilde{\Omega}_{2}G\\
i\gamma\tilde{\Omega}_{2}^{\ast}G & 1-i\gamma G_{2}%
\end{array}
\right]  \left[
\begin{array}
[c]{c}%
\tilde{R}_{31}\\
\tilde{R}_{21}%
\end{array}
\right]  =\left[
\begin{array}
[c]{c}%
-G_{1}\\
\Omega_{2}^{\ast}G
\end{array}
\right]  \tilde{\Omega}_{1}(\mathbf{k},\omega)n_{0},\label{url-1}%
\end{equation}
where we have introduced the integrals
\begin{subequations}
\label{gd}%
\begin{align}
G(\mathbf{k},\omega) &  =\int d^{3}vF\left(  \mathbf{v}\right)  \frac{1}%
{\xi_{d}},\label{gd-1}\\
G_{1}(\mathbf{k},\omega) &  =\int d^{3}vF\left(  \mathbf{v}\right)
\frac{\omega-\mathbf{k\cdot v}+\xi_{2}}{\xi_{d}},\label{gd-2}\\
G_{2}\left(  \mathbf{k},\omega\right)   &  =\int d^{3}vF\left(  \mathbf{v}%
\right)  \frac{\omega-\mathbf{k\cdot v}+\xi_{1}}{\xi_{d}}.\label{gd-3}%
\end{align}
Solving Eq.(\ref{url-1}) for $\tilde{R}_{31}$ we obtain
\end{subequations}
\begin{equation}
\tilde{R}_{31}\left(  \mathbf{k},\omega\right)  =i\tilde{\Omega}_{1}\left(
\mathbf{k},\omega\right)  \frac{n_{0}}{\gamma}\left(  \frac{1-i\gamma
G_{2}\left(  \mathbf{k},\omega\right)  }{G_{d}\left(  \mathbf{k}%
,\omega\right)  }-1\right)  ,\label{url-2}%
\end{equation}
where%
\begin{equation}
G_{d}\left(  \mathbf{k},\omega\right)  =\left(  1-i\gamma G_{1}\right)
\left(  1-i\gamma G_{2}\right)  +\gamma^{2}|\Omega_{2}|^{2}G^{2}.\label{Dg-1}%
\end{equation}

We can now return to Eq.(\ref{Ert-5}) and solve for $\tilde{\Omega}_{1}\left(
\mathbf{k},\omega\right)  $. Since $\tilde{R}_{31}$ is linear in
$\tilde{\Omega}_{1}$, we introduce the linear susceptibility by,%
\begin{equation}
\tilde{R}_{31}\left(  \mathbf{k},\omega\right)  =\chi_{31}\left(
\mathbf{k},\omega\right)  \frac{c}{g}\tilde{\Omega}_{1}\left(  \mathbf{k}%
,\omega\right)  ,\label{chi}%
\end{equation}
and the complex wave-number,%
\begin{equation}
p\left(  \mathbf{k},\omega\right)  =\frac{\omega}{c}-\frac{k^{2}}{2q_{1}}%
+\chi_{31}\left(  \mathbf{k},\omega\right)  ,\label{pk-1}%
\end{equation}
to express Eq.(\ref{Ert-5}) as%
\begin{equation}
\left[  k_{z}-p\left(  \mathbf{k},\omega\right)  \right]  \tilde{\Omega}%
_{1}(\mathbf{k},\omega)=0.\label{Ert-6}%
\end{equation}
When the changes in the envelopes along the $z-$direction are much smaller
compared to the changes in the transverse plane, we may replace
$\mathbf{k\rightarrow k}_{\perp}$ in the $\xi$'s and the $G$'s in
Eqs.(\ref{D_d-1}) and (\ref{gd}), where $\mathbf{k}_{\perp}$ is the projection
of $\mathbf{k}$ onto the transverse plane, and write Eq.(\ref{Ert-6}) in the
$(z;\mathbf{k}_{\perp},\omega)$ coordinates:%
\begin{equation}
\frac{\partial}{\partial z}\tilde{\Omega}_{1}\left(  z;\mathbf{k}_{\perp
},\omega\right)  =ip\left(  \mathbf{k}_{\perp},\omega\right)  \tilde{\Omega
}_{1}(z;\mathbf{k}_{\perp},\omega),\label{Ert-7}%
\end{equation}
or, the solution%
\begin{equation}
\tilde{\Omega}_{1}\left(  z_{2};\mathbf{k}_{\perp},\omega\right)
=\tilde{\Omega}_{1}\left(  z_{1};\mathbf{k}_{\perp},\omega\right)
e^{ip\left(  \mathbf{k}_{\perp},\omega\right)  \left(  z_{2}-z_{1}\right)
}.\label{Ert-8}%
\end{equation}
Finally, the probe's envelope is given by%
\begin{align}
\tilde{\Omega}_{1}\left(  z_{2};x,y,t\right)   & =\int_{-\infty}^{\infty}%
\frac{d\omega}{2\pi}e^{-i\omega t}\int_{-\infty}^{\infty}\frac{d^{2}k_{\perp}%
}{2\pi}e^{ik_{x}x+ik_{y}y}\nonumber\\
& \times\tilde{\Omega}_{1}\left(  z_{1};\mathbf{k}_{\perp},\omega\right)
e^{ip\left(  \mathbf{k}_{\perp},\omega\right)  \left(  z_{2}-z_{1}\right)
}.\label{ea-1}%
\end{align}

\subsection*{The Doppler-Dicke Transition}

It is first instructive to consider the case of ordinary one-photon absorption
by taking $\Omega_{2}=0$, and to assume for simplicity a plane-wave stationary
probe, namely $\omega=0$ and $\mathbf{k}=0$. For this case Eqs.(\ref{gd}%
)-(\ref{chi}) give
\begin{equation}
\chi_{31}\left(  \Delta_{1}\right)  =i\frac{gn_{0}}{c}K\left(  \Delta
_{1}\right)  ,\label{chi_1p}%
\end{equation}
where $K\left(  \Delta_{1}\right)  $ is the one-photon complex spectrum,%
\begin{equation}
K\left(  \Delta_{1}\right)  =\frac{iG_{1}\left(  \Delta_{1}\right)
}{1-i\gamma G_{1}\left(  \Delta_{1}\right)  },\label{K}%
\end{equation}
and $G_{1}\left(  \Delta_{1}\right)  $ takes the form of a Doppler-like
profile (a Voigt convolution), with $\gamma$ being added to the standard
homogenous-width, $\Gamma_{d}$:
\begin{equation}
G_{1}\left(  \Delta_{1}\right)  =\frac{1}{\sqrt{2\pi}v_{\text{th}}}\int
du\frac{e^{-u^{2}/(2v_{\text{th}}^{2})}}{\Delta_{1}-q_{1}u+i\left(  \Gamma
_{d}+\gamma\right)  }.\label{g1}%
\end{equation}
The spectrum $K(\Delta_{1})$ in the form of Eq.(\ref{K}) was previously
presented by May \cite{May1999} for one-photon transitions. Its extreme limits
are \cite{GalatryPR1961,May1999,FirstenbergPRA2007}: the \emph{Doppler limit},
trivially obtained by setting $\gamma=0,$ and the \emph{Dicke limit}, found
for large $\gamma$. The Dicke parameter is $v_{\text{th}}q_{1}/\gamma$,
proportional to the ratio between the mean free-path and the radiation wavelength.

For optical transitions in room-temperature vapor, the one-photon line is
usually in the far Doppler limit, i.e. $K\approx iG_{1}$. We have defined $K$
such that it is real for $\left\vert \Delta_{1}\right\vert \ll\left\vert
q_{1}u+i\left(  \Gamma_{d}+\gamma\right)  \right\vert ,$ i.e. near the
one-photon resonance where experiments in EIT are often done, and is equal to
the on-resonance absorption (in frequency units). Specifically, for an atom at
rest ($\gamma=0$ and no Doppler), $K\left(  \Delta_{1}=0\right)  =\Gamma
_{d}^{-1}$.%

%TCIMACRO{\FRAME{ftbpFU}{8.6042cm}{6.4603cm}{0pt}{\Qcb{Normalized EIT
%transmission spectra, numerically calculated from Eqs. (\ref{gd})-(\ref{chi}),
%for three values of the EIT Dicke parameter, $\eta=v_{th}k/\gamma$ (the ratio
%between the residual Doppler width and the velocity relaxation rate), with
%$\Gamma_{d}=2500v_{th}k$, $\Gamma_{21}=0.025v_{th}k$ and $\left\vert
%\Omega_{2}\right\vert ^{2}/\Gamma_{d}=\Gamma_{21}/25$ (small
%power-broadening). When $\eta$ is large (solid black line) the spectrum is a
%Voigt curve (a Gaussian-Lorentzian convolution). When $\eta$ is small (dashed
%blue line) the spectrum is a pure Lorentzian of width $\Gamma_{d}$. The
%dot-dashed green line demonstrates an intermediate result.}}{\Qlb{fig_chi_2p}%
%}{chi_2photon.eps}{\special{ language "Scientific Word";  type "GRAPHIC";
%width 8.6042cm;  height 6.4603cm;  depth 0pt;  original-width 14.8184cm;
%original-height 11.1017cm;  cropleft "0";  croptop "1";  cropright "1";
%cropbottom "0";  filename 'Chi_2photon.eps';file-properties "XNPEU";}}}%
%BeginExpansion
\begin{figure}
[ptb]
\begin{center}
\includegraphics[
height=6.4603cm,
width=8.6042cm
]%
{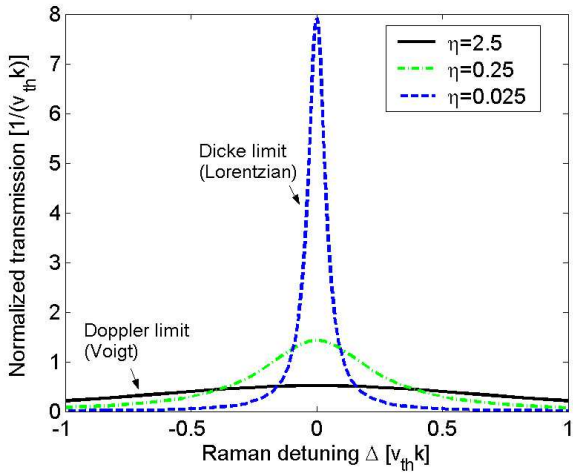}%
\caption{Normalized EIT transmission spectra, numerically calculated from Eqs.
(\ref{gd})-(\ref{chi}), for three values of the EIT Dicke parameter,
$\eta=v_{th}k/\gamma$ (the ratio between the residual Doppler width and the
velocity relaxation rate), with $\Gamma_{d}=2500v_{th}k$, $\Gamma
_{21}=0.025v_{th}k$ and $\left\vert \Omega_{2}\right\vert ^{2}/\Gamma
_{d}=\Gamma_{21}/25$ (small power-broadening). When $\eta$ is large (solid
black line) the spectrum is a Voigt curve (a Gaussian-Lorentzian convolution).
When $\eta$ is small (dashed blue line) the spectrum is a pure Lorentzian of
width $\Gamma_{d}$. The dot-dashed green line demonstrates an intermediate
result.}%
\label{fig_chi_2p}%
\end{center}
\end{figure}
%EndExpansion

\begin{figure}
[ptb]
\begin{center}
\includegraphics[
trim=0.000000in 0.000000in 0.264864in 0.000000in,
height=6.6821cm,
width=8.5009cm
]%
{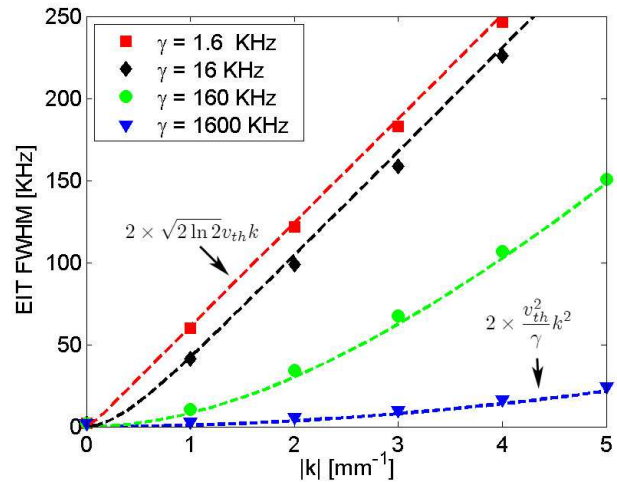}%
\caption{Full width at half maximum of the EIT transparency window (points),
obtained from numerical results similar to Fig. \ref{fig_chi_2p}, as a
function of the wave-vector difference, $k,$ for various values of the
velocity relaxation rate, $\gamma$. The dashed lines are given by
Eq.(\ref{eq_analytic_H}). Other parameters are (typical for experiments with
small power-broadening): $v_{th}=170$ m/s, $\Gamma_{d}=100$ MHz, $\Gamma
_{21}=1$ KHz, $|\Omega_{2}|^{2}/\Gamma_{d}=40$ Hz. The three lines in Fig.
\ref{fig_chi_2p} correspond here to $\left\vert k\right\vert \approx1.5$
mm$^{-1}$ and $\gamma=16,160$ and $1600$ KHz.}%
\label{fig_FWHM_2p}%
\end{center}
\end{figure}

A similar Doppler-Dicke transition occurs for the EIT transmission window. In
order to demonstrate that, we have chosen a set of typical parameters with
small power-broadening and calculated several EIT lines from Eqs.(\ref{gd}%
)-(\ref{chi}) by numerically integrating Eqs.(\ref{gd}). Three line shapes as
a function of the normalized Raman detuning are presented in Figs.
\ref{fig_chi_2p}. The full-width at half-maximum (FWHM), as a function of
$|\mathbf{k|}$ and for various values of $\gamma$, is presented in Fig.
\ref{fig_FWHM_2p}. For the calculations we took a stationary ($\omega=0$),
collinear and degenerate ($\mathbf{q}_{1}=\mathbf{q}_{2}$) EIT with
$\mathbf{k\perp\hat{z}}$, so that $k=\left\vert \mathbf{k}\right\vert $ is the
wave-vector difference between the pump and the probe (similar results are
obtained by replacing $k$ with $|\mathbf{q}_{1}-\mathbf{q}_{2}|,$ when
$|\mathbf{q}_{1}-\mathbf{q}_{2}|\ll q_{1}$). The residual Doppler width is
expected to be $v_{\text{th}}k,$ and the EIT-Dicke parameter is $\eta
=v_{\text{th}}k/\gamma$ \cite{FirstenbergPRA2007}. Fig. \ref{fig_FWHM_2p}
clearly shows the transition between the linear (Doppler) regime, where
$\eta\gg1$, to the quadratic (Dicke) regime, where $\eta\ll1$. The results of
the numerical integrations throughout the Doppler-Dicke transition are well
approximated (dashed lines in Fig. \ref{fig_FWHM_2p}) by the analytic
expression:%
\begin{equation}
\text{FWHM}=2\times\frac{2}{a^{2}}\gamma H\left(  a\frac{v_{\text{th}}%
k}{\gamma}\right)  ,\label{eq_analytic_H}%
\end{equation}
where $H\left(  x\right)  =e^{-x}-1+x$ and $a^{2}=2/\ln2$. The function
$H\left(  x\right)  $ is usually associated with the velocity self-correlation
in Brownian motion \cite{FirstenbergPRA2007,ChandrasekharRMP1943}, and its
extreme limits are $H(x\rightarrow0)=x^{2}/2$ and $H(x\rightarrow\infty)=x $.%

%TCIMACRO{\FRAME{ftbpFU}{8.5009cm}{6.6821cm}{0pt}{\Qcb{Full width at half
%maximum of the EIT transparency window (points), obtained from numerical
%results similar to Fig. \ref{fig_chi_2p}, as a function of the wave-vector
%difference, $k,$ for various values of the velocity relaxation rate, $\gamma$.
%The dashed lines are given by Eq.(\ref{eq_analytic_H}). Other parameters are
%(typical for experiments with small power-broadening): $v_{th}=170$ m/s,
%$\Gamma_{d}=100$ MHz, $\Gamma_{21}=1$ KHz, $|\Omega_{2}|^{2}/\Gamma_{d}=40$
%Hz. The three lines in Fig. \ref{fig_chi_2p} correspond here to $\left\vert
%k\right\vert \approx1.5$ mm$^{-1}$ and $\gamma=16,160$ and $1600$ KHz.}%
%}{\Qlb{fig_FWHM_2p}}{fwhm_2p.eps}{\special{ language "Scientific Word";
%type "GRAPHIC";  maintain-aspect-ratio TRUE;  display "USEDEF";
%valid_file "F";  width 8.5009cm;  height 6.6821cm;  depth 0pt;
%original-width 5.834in;  original-height 4.3708in;  cropleft "0";
%croptop "1";  cropright "0.9546";  cropbottom "0";
%filename 'FWHM_2p.eps';file-properties "XNPEU";}}}%
%BeginExpansion
%EndExpansion

\section{The Dicke-Diffusion Solution\label{ch_dicke_diff}}

In most realistic experiments with EIT, $\gamma$ is large enough to cause
Dicke-narrowing of the two-photon line. In principle, this spectrum can be
obtained by applying the large $\gamma$ limit to the general solution of the
susceptibility of the system, Eqs. (\ref{url-2}) and (\ref{chi}). However,
since Eq.(\ref{url-2}) is somewhat opaque, it seems worthwhile to first derive
a diffusion-like equation for $R_{21}(\mathbf{r},t),$ and then find the
response of $R_{31}(\mathbf{r},t).$ We do this first in the absence of fields
and then in their presence, analyzing the cases of stationary and
non-stationary probe.

\subsection{Diffusion during storage-of-light\label{ss_storage}}

We are interested in describing the dynamics of the ground-state populations
and coherences,
\begin{equation}
R_{G}=\left(
\begin{array}
[c]{cc}%
R_{11} & R_{21}^{\ast}\\
R_{21} & R_{22}%
\end{array}
\right)  ,
\end{equation}
in the absence of electromagnetic fields. This amounts to the situation
investigated in recent storage-of-light experiments
\cite{PugatchPRL2007,ShukerImaging2007}, in which a probe beam with a
non-trivial spatial envelope in the transverse plane was stored in an EIT
vapor and then retrieved. Following Ref. \cite{LukinPRL2000}, the storage
procedure was described by a linear mapping of the probe field onto the
ground-state coherence, $R_{21}\left(  \mathbf{r},t\right)  $. It was then
presumed that the dynamics "in the dark" can be described by the standard
diffusion process, namely $[\dot{R}_{G}]_{\text{diffusion}}=D\nabla^{2}R_{G},$
where $D$ is the diffusion coefficient \cite{PugatchPRL2007}. Eventually the
retrieved probe field is obtained from $R_{21}\left(  \mathbf{r},t\right)  $
by the reverse linear mapping.

The dynamics of $R_{21}\left(  \mathbf{r},t\right)  $ can be described in
terms of a diffusion-like equation. To this end we return to
Eqs.(\ref{ul-rl-3}), substitute $\Omega_{1}=\Omega_{2}=0$, and apply a
procedure similar to that of Chapman-Enskog (see e.g. \cite{ChapmanBook}). The
equation for $\rho_{21}$ is%

\begin{equation}
\left[  \frac{\partial}{\partial t}+\Gamma_{21}+\gamma+i\omega_{21}%
+\mathbf{v\cdot}\frac{\partial}{\partial\mathbf{r}}\right]  \rho_{21}=\gamma
R_{21}(\mathbf{r},t)F\left(  \mathbf{v}\right)  .\label{ab1-storage}%
\end{equation}
Integrating over velocity and using Eqs.(\ref{rs-4}) and (\ref{Fv}), we have%
\begin{equation}
\left[  \frac{\partial}{\partial t}+i\left(  \omega_{21}-i\Gamma_{21}\right)
\right]  R_{21}(\mathbf{r},t)+\frac{\partial}{\partial\mathbf{r}}\mathbf{\cdot
J}_{21}(\mathbf{r},t)=0,\label{ab2-storage}%
\end{equation}
where%
\begin{equation}
\mathbf{J}_{ss^{\prime}}(\mathbf{r},t)=\int d^{3}v\mathbf{v}\rho_{ss^{\prime}%
}(\mathbf{r},\mathbf{v},t)\label{Jss}%
\end{equation}
is the current density of the density-matrix. When $\gamma$\ is dominant, we
divide Eq.(\ref{ab1-storage}) by $\gamma$ and get, to zero order in
$1/\gamma,$
\begin{equation}
\rho_{21}^{\left(  0\right)  }(\mathbf{r},\mathbf{v},t)=R_{21}(\mathbf{r}%
,t)F\left(  \mathbf{v}\right)  ,
\end{equation}
which carries no current. Thus, to find the current we go to first order in
$1/\gamma$,
\begin{equation}
\rho_{21}(\mathbf{r},\mathbf{v},t)=R_{21}(\mathbf{r},t)F\left(  \mathbf{v}%
\right)  +\frac{1}{\gamma}\rho_{21}^{\left(  1\right)  }(\mathbf{r}%
,\mathbf{v},t),\label{aa-5-storage}%
\end{equation}
substitute it into Eq.(\ref{ab1-storage}), multiply by $v_{j}$ and integrate
over velocity,%
\begin{align}
\int d^{3}vv_{j}\left[  v_{i}\frac{\partial}{\partial x_{i}}+\frac{\partial
}{\partial t}+\Gamma_{21}+\gamma+i\omega_{21}\right]   & \label{ad-storage}\\
\times\left(  R_{21}(\mathbf{r},t)F\left(  \mathbf{v}\right)  +\frac{1}%
{\gamma}\rho_{21}^{\left(  1\right)  }(\mathbf{r},\mathbf{v},t)\right)   &
=0.\nonumber
\end{align}
Since%
\begin{equation}
\int d^{3}vv_{j}v_{i}\frac{\partial}{\partial x_{i}}R_{21}(\mathbf{r}%
,t)F\left(  \mathbf{v}\right)  =\delta_{ij}v_{\text{th}}^{2}\frac{\partial
}{\partial x_{j}}R_{21}(\mathbf{r},t),\label{ad-2}%
\end{equation}
and%
\begin{equation}
\int d^{3}vv_{j}\rho_{21}^{\left(  1\right)  }(\mathbf{r},\mathbf{v},t)=\gamma
J_{21,j}(\mathbf{r},t),\label{ad-3}%
\end{equation}
we find from Eq.(\ref{ad-storage}), keeping only leading terms in $\gamma,$%
\begin{equation}
\mathbf{J}_{21}(\mathbf{r},t)=-D\frac{\partial}{\partial\mathbf{r}}%
R_{21}(\mathbf{r},t),
\end{equation}
where%
\begin{equation}
D=\frac{v_{\text{th}}^{2}}{\gamma}\label{dd-1}%
\end{equation}
is the spatial diffusion coefficient \cite{ChandrasekharRMP1943}. Substituting
this result into Eq.(\ref{ab2-storage}), we find%
\begin{equation}
\frac{\partial}{\partial t}R_{21}\left(  \mathbf{r},t\right)  =D\nabla
^{2}R_{21}(\mathbf{r},t)-\left(  \Gamma_{21}+i\omega_{21}\right)
R_{21}(\mathbf{r},t).\label{R21-diff}%
\end{equation}
Eq.(\ref{R21-diff}) describes a spatial diffusion of the coherence,
accompanied by a homogenous decay of rate $\Gamma_{21}$ and a rotation of rate
$\omega_{21}$. A similar derivation can be preformed for the ground-state
populations, $R_{11}$ and $R_{22}$, and it results in a similar diffusion
equation. This solution affirms the theoretical conjectures of Refs.
\cite{PugatchPRL2007,ShukerImaging2007}. In what follows, it is generalized to
describe the diffusion in the presence of the fields, i.e. during slow-light propagation.

It is interesting to note that even in the limit $\Gamma_{21}\rightarrow0$,
Eq.(\ref{R21-diff}) results in the decay of the total stored-light energy.
Assuming the stored coherence, $R_{21}$, is linear in the field's amplitude,
the total intensity is proportional to the integral over $|R_{21}|^{2}$, which
is not conserved and always decreases under diffusive spread. A similar
observation was made in \cite{Zimmer2006} in the context of stationary light
pulses that diffuse along the $z-$direction (the diffusion equation therein
originates from a different mechanism).

\subsection{Diffusion in the presence of fields}

Here we derive the dynamic equations for the envelopes of the densities,
$\tilde{R}_{21}\left(  \mathbf{r},t\right)  $ and $\tilde{R}_{31}\left(
\mathbf{r},t\right)  ,$ along similar lines as above, while considering the
interaction with the pump and the probe. For brevity, we omit the $\left(
\mathbf{r},t\right)  $ notation and denote $\delta\mathbf{q=q}_{1}%
-\mathbf{q}_{2}$. We start from the envelope equations, Eqs.(\ref{ul-rl-4}),
and integrate Eq.(\ref{rl-4}) over velocity,
\begin{equation}
\left(  \frac{\partial}{\partial\mathbf{r}}+i\delta\mathbf{q}\right)
\mathbf{\cdot\tilde{J}}_{21}+\left(  \frac{\partial}{\partial t}%
-i\Delta+\Gamma_{21}\right)  \tilde{R}_{21}=i\tilde{\Omega}_{2}^{\ast
}(\mathbf{r})\tilde{R}_{31},\label{aa-4}%
\end{equation}
where $\mathbf{\tilde{J}}_{ss^{\prime}}$ is the envelope of the current
densities, defined in analogy to Eq.(\ref{Jss}). We expand $\tilde{\rho}_{21}
$ as in Eq.(\ref{aa-5-storage}), $\tilde{\rho}_{21}=\tilde{R}_{21}F\left(
\mathbf{v}\right)  +\left(  1/\gamma\right)  \tilde{\rho}_{21}^{\left(
1\right)  },$ multiply Eq.(\ref{rl-4}) by $\mathbf{v}$ and integrate over
velocity. Using Eqs.(\ref{ad-2}) and (\ref{ad-3}), and keeping leading terms
in $\gamma,$ we find%
\begin{equation}
\mathbf{\tilde{J}}_{21}+D\left(  \frac{\partial}{\partial\mathbf{r}}%
+i\delta\mathbf{q}\right)  \tilde{R}_{21}=i\frac{\tilde{\Omega}_{2}^{\ast
}(\mathbf{r})}{\gamma}\mathbf{\tilde{J}}_{31},\label{aa-6}%
\end{equation}
where $D$ is defined in Eq.(\ref{dd-1}). Substituting $\mathbf{\tilde{J}}%
_{21}$ back into Eq.(\ref{aa-4}) we get
\begin{align}
& \left[  \frac{\partial}{\partial t}-i\Delta+\Gamma_{21}-D\left(
\frac{\partial}{\partial\mathbf{r}}+i\delta\mathbf{q}\right)  ^{2}\right]
\tilde{R}_{21}\nonumber\\
& =i\tilde{\Omega}_{2}^{\ast}(\mathbf{r})\tilde{R}_{31}-i\left(
\frac{\partial}{\partial\mathbf{r}}+i\delta\mathbf{q}\right)  \mathbf{\cdot
}\frac{\tilde{\Omega}_{2}^{\ast}(\mathbf{r})}{\gamma}\mathbf{\tilde{J}}%
_{31}.\label{aa-8}%
\end{align}

In order to calculate $\tilde{R}_{31}$ and $\mathbf{\tilde{J}}_{31}$, we
assume in Eq.(\ref{ul-4}) that temporal and spatial changes in the envelope of
the probe are much smaller then the one-photon homogenous decoherence rate
$\left(  \Gamma_{d}+\gamma\right)  $ and the wave-number $\left(
q_{1}\right)  ,$ respectively:%
\begin{equation}
\left\vert \frac{\partial}{\partial t}+\mathbf{v\cdot}\frac{\partial}%
{\partial\mathbf{r}}\right\vert \ll\left\vert \xi_{1}\right\vert =\left\vert
\Delta_{1}-\mathbf{q}_{1}\mathbf{\cdot v}+i\left(  \Gamma_{d}+\gamma\right)
\right\vert .\label{vr-2}%
\end{equation}
We then formally solve Eq.(\ref{ul-4}) for $\tilde{\rho}_{31}$ and substitute
\emph{only the dominant part} of $\tilde{\rho}_{21},$ i.e. $\tilde{\rho}%
_{21}^{\left(  0\right)  }=\tilde{R}_{21}F\left(  \mathbf{v}\right)  $, to
find%
\begin{equation}
\tilde{\rho}_{31}(\mathbf{r},\mathbf{v},t)=\frac{\left[  i\gamma\tilde{R}%
_{31}-\tilde{\Omega}_{2}(\mathbf{r})\tilde{R}_{21}-\tilde{\Omega}%
_{1}(\mathbf{r},t)n_{0}\right]  F\left(  \mathbf{v}\right)  }{\left(
\Delta_{1}-\mathbf{q}_{1}\mathbf{\cdot v}+i\left(  \Gamma_{d}+\gamma\right)
\right)  }.\label{aa-9}%
\end{equation}
Integrating Eq.(\ref{aa-9}) over velocity, we get%
\begin{equation}
\tilde{R}_{31}=G_{1}\left[  i\gamma\tilde{R}_{31}-\tilde{\Omega}%
_{2}(\mathbf{r})\tilde{R}_{21}-\tilde{\Omega}_{1}(\mathbf{r},t)n_{0}\right]
,\label{ac-1}%
\end{equation}
where $G_{1}$ of Eq.(\ref{g1}) is the Doppler profile, or%
\begin{equation}
\tilde{R}_{31}(\mathbf{r},t)=iK\left[  \tilde{\Omega}_{1}(\mathbf{r}%
,t)n_{0}+\tilde{\Omega}_{2}(\mathbf{r})\tilde{R}_{21}(\mathbf{r},t)\right]
,\label{ac-2}%
\end{equation}
where $K=K\left(  \Delta_{1}\right)  $ of Eq.(\ref{K}) is the one-photon
absorption spectrum, which in the Dicke limit of the EIT can be considered as
a constant near the EIT line.

Eq.(\ref{aa-8}) and Eq.(\ref{ac-2}) form a complete set for $\tilde{R}_{21}$
and $\tilde{R}_{31},$ only when the term
\begin{equation}
\left(  \frac{\partial}{\partial\mathbf{r}}+i\delta\mathbf{q}\right)
\mathbf{\cdot}\frac{\tilde{\Omega}_{2}^{\ast}(\mathbf{r})}{\gamma
}\mathbf{\tilde{J}}_{31}\label{term}%
\end{equation}
in Eq.(\ref{aa-8}) can be neglected. This term vanishes completely in the
special case of pump and probe which are plane-waves ($\partial/\partial
\mathbf{r=}0$), collinear and degenerate ($\delta\mathbf{q}=0$). It can also
be neglected whenever $|\Omega_{2}|\ll\gamma,$ as is the case in many
realistic situations, cf. \cite{ShukerPRA2007}. Furthermore, if the latter
condition is not satisfied, we can still neglect the term (\ref{term}) when
both the spatial variations ($\partial/\partial\mathbf{r}$) and $\delta
\mathbf{q}$ reside in the transverse plane, perpendicular to $\mathbf{q}_{1}$,
since $\mathbf{\tilde{J}}_{31}\parallel\mathbf{q}_{1}$ (as can be found by
multiplying Eq.(\ref{aa-9}) by $\mathbf{v}$ and integrating). Discarding this
term from Eq.(\ref{aa-8}), and together with Eq.(\ref{ac-2}), we find for
$\tilde{R}_{21}$ a diffusion-like equation,
\begin{align}
& \left[  \frac{\partial}{\partial t}-i\Delta+\Gamma_{21}+K|\tilde{\Omega}%
_{2}(\mathbf{r})|^{2}\right]  \tilde{R}_{21}\left(  \mathbf{r},t\right)
\label{ac-6}\\
& =D\left(  \frac{\partial}{\partial\mathbf{r}}+i\delta\mathbf{q}\right)
^{2}\tilde{R}_{21}\left(  \mathbf{r},t\right)  -n_{0}K\tilde{\Omega}_{2}%
^{\ast}(\mathbf{r})\tilde{\Omega}_{1}(\mathbf{r},t).\nonumber
\end{align}
Eq.(\ref{ac-6}) is the extension of Eq.(\ref{R21-diff}) in the presence of
fields, and it is written in terms of the envelopes. The term $i\delta
\mathbf{q}$ is responsible for the diffusion across the fields' interference
pattern, created when $\mathbf{q}_{1}\neq\mathbf{q}_{2}.$ Once we solve the
diffusion equation for $\tilde{R}_{21}(\mathbf{r},t)$ we substitute it in
Eq.(\ref{ac-2}) and obtain $\tilde{R}_{31}(\mathbf{r},t).$ We carry this out
in the next subsection for the case of a plane-wave pump and in section
\ref{ch_ramsey} for a finite-size pump.

\subsection{The Dicke-like Absorption Spectrum}

We would like to calculate the susceptibility of the system for the case of a
plane-wave pump, $\tilde{\Omega}_{2}(\mathbf{r})=\Omega_{2}$. We Fourier
transform in $\mathbf{r}$ and $t$ using Eq.(\ref{ft-1}) and turn
Eq.(\ref{ac-6}) into%
\begin{align}
& \left[  i\left(  \Delta+\omega\right)  -\Gamma_{21}-K|\Omega_{2}%
|^{2}\right]  \tilde{R}_{21}\left(  \mathbf{k},\omega\right) \label{ac-8}\\
& =D\left(  \delta\mathbf{q}+\mathbf{k}\right)  ^{2}\tilde{R}_{21}\left(
\mathbf{k},\omega\right)  +n_{0}K\Omega_{2}^{\ast}\tilde{\Omega}%
_{1}(\mathbf{k},\omega),\nonumber
\end{align}
and Eq.(\ref{ac-2}) into%
\begin{equation}
\tilde{R}_{31}\left(  \mathbf{k},\omega\right)  =iK\left[  \tilde{\Omega}%
_{1}\left(  \mathbf{k},\omega\right)  n_{0}+\Omega_{2}\tilde{R}_{21}\left(
\mathbf{k},\omega\right)  \right]  .\label{ac-7}%
\end{equation}
Solving Eq.(\ref{ac-8}) for $\tilde{R}_{21}\left(  \mathbf{k},\omega\right)
$, substituting in Eq.(\ref{ac-7}) to obtain $\tilde{R}_{31}$ and using
Eq.(\ref{chi}), we find the susceptibility in the diffusion-limit case to be%
\begin{equation}
\chi_{31}\left(  \mathbf{k},\omega\right)  =\frac{g}{c}iKn_{0}\left[
1-L\left(  \mathbf{k},\omega\right)  \right]  ,\label{chi-2}%
\end{equation}
where%
\begin{equation}
L\left(  \mathbf{k},\omega\right)  =\frac{-K|\Omega_{2}|^{2}}{i\left(
\Delta+\omega\right)  -\Gamma_{\hom}-D\left(  \delta\mathbf{q}+\mathbf{k}%
\right)  ^{2}},\label{Lkw}%
\end{equation}
and%
\begin{equation}
\Gamma_{\hom}=\Gamma_{21}+K|\Omega_{2}|^{2}%
\end{equation}
is the EIT width in the absence of diffusion.

The absorption of the probe is proportional to $\Im m\chi_{31}$. As explained
in section \ref{ch_general_solution}, in typical cases $K$ is real and
therefore%
\begin{equation}
\Im m\chi_{31}=\frac{g}{c}Kn_{0}\left[  1-\Re eL\left(  \mathbf{k}%
,\omega\right)  \right]  ,\label{ImChi-2}%
\end{equation}
i.e. the well-known EIT absorption spectrum is composed of the one-photon
absorption, $gKn_{0}/c$, and a "transparency window", $\Re eL\left(
\mathbf{k},\omega\right)  $, of Lorentzian shape:%
\begin{equation}
\Re eL\left(  \mathbf{k},\omega\right)  =\frac{K|\Omega_{2}|^{2}[\Gamma_{\hom
}+D\left(  \delta\mathbf{q}+\mathbf{k}\right)  ^{2}]}{\left(  \Delta
+\omega\right)  ^{2}+[\Gamma_{\hom}+D\left(  \delta\mathbf{q}+\mathbf{k}%
\right)  ^{2}]^{2}}.\label{ReL}%
\end{equation}
Considering $L\left(  \mathbf{k},\omega\right)  $ for a given $\mathbf{k}$, as
a function of the Raman detuning $\left(  \Delta+\omega\right)  $, we find the
homogenous EIT width to be $\Gamma_{\hom}=\Gamma_{21}+K|\Omega_{2}|^{2},$
broadened by the Dicke-EIT width, $D\left(  \mathbf{q}_{1}-\mathbf{q}%
_{2}+\mathbf{k}\right)  ^{2}$. Eqs.(\ref{ImChi-2})\ and (\ref{ReL}) generalize
the results of our previous work in Ref.\cite{FirstenbergPRA2007} for a finite
probe in space and time. It includes the \emph{power-broadening} effect, $K$%
%TCIMACRO{\TEXTsymbol{\vert}}%
%BeginExpansion
$\vert$%
%EndExpansion
$\Omega_{2}|^{2}$ that was absent in Ref.\cite{FirstenbergPRA2007}, in which
the low-contrast approximation was taken. Notice that the term $K|\Omega
_{2}|^{2}$ replaces the standard power-broadening term, $|\Omega_{2}%
|^{2}/\Gamma_{d}$, to incorporate the Doppler-broadening of the one-photon line.

\subsection{Spatial-frequency filter and diffusion-like behavior}%

%TCIMACRO{\FRAME{ftbpFU}{8.222cm}{6.1791cm}{0pt}{\Qcb{The spatial-frequency
%filter (normalized EIT transmission), given in Eq.(\ref{ReL}), as a function
%of $\left\vert \mathbf{k}_{\perp}\right\vert =\left\vert k\mathbf{\hat{x}%
%}\right\vert \mathbf{,}$ for different Raman detunings, with $\omega=0,$
%$\delta\mathbf{q}=0$ and$\ \Gamma_{21}=K|\Omega_{2}|^{2}$. The solid-black
%curve is plotted for $\Delta=0$. The dashed red and blue curves demonstrate
%the effect of non-zero Raman detuning: a decrease in transparency alongside a
%change in the curvature near $k=0$.}}{\Qlb{fig_relk}}{rel2.eps}%
%width 8.222cm;  height 6.1791cm;  depth 0pt;  original-width 14.8184cm;
%original-height 11.1017cm;  cropleft "0";  croptop "1";  cropright "1";
%cropbottom "0";  filename 'ReL2.eps';file-properties "XNPEU";}}}%
%BeginExpansion
\begin{figure}
[ptb]
\begin{center}
\includegraphics[
height=6.1791cm,
width=8.222cm
]%
{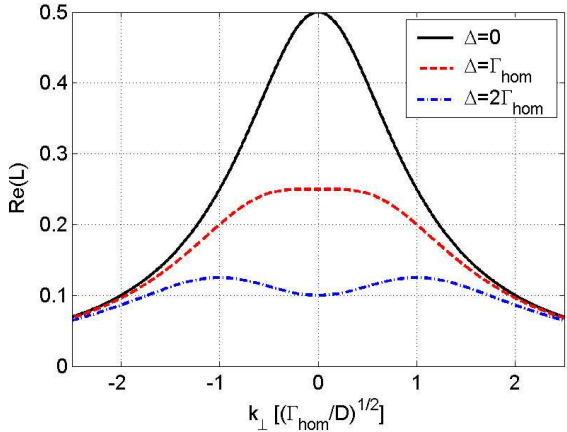}%
\caption{The spatial-frequency filter (normalized EIT transmission), given in
Eq.(\ref{ReL}), as a function of $\left\vert \mathbf{k}_{\perp}\right\vert
=\left\vert k\mathbf{\hat{x}}\right\vert \mathbf{,}$ for different Raman
detunings, with $\omega=0,$ $\delta\mathbf{q}=0$ and$\ \Gamma_{21}%
=K|\Omega_{2}|^{2}$. The solid-black curve is plotted for $\Delta=0$. The
dashed red and blue curves demonstrate the effect of non-zero Raman detuning:
a decrease in transparency alongside a change in the curvature near $k=0$.}%
\label{fig_relk}%
\end{center}
\end{figure}
%EndExpansion

When $\Im m\chi_{31}\left(  \mathbf{k},\omega\right)  $ is considered in
steady-state ($\omega=0$) as a function of $\mathbf{k}$, it acts as a
\emph{spatial-frequency filter} for the probe beam. We have in mind a
stationary probe beam in the plane $z=z_{1}$ with an envelope $\tilde{\Omega
}_{1}\left(  z_{1};x,y\right)  $, which propagates through the medium to the
$z=z_{2}$ plane. Following Eq.(\ref{ea-1}), $\Im m\chi_{31}\left(
\mathbf{k}_{\bot},\omega\right)  $ determines the absorption of each
spatial-frequency component of $\tilde{\Omega}_{1}\left(  z;\mathbf{k}_{\bot
}\right)  $. Since the first term in Eq.(\ref{ImChi-2}) -- the one-photon
absorption -- is constant for all $\mathbf{k}_{\bot},$ the filter becomes a
"transmission" filter with the shape $\Re eL\left(  \mathbf{k}_{\bot}\right)
.$ Several examples for this filter are plotted in Fig. \ref{fig_relk} with
$\Gamma_{\hom}=2\Gamma_{12}$ and $\delta\mathbf{q}=0$. On Raman-resonance, the
curve is a Lorentzian of width $k_{\text{typ}}=\sqrt{\Gamma_{\hom}/D},$ and
the maximum overall transmission is achieved. For non-zero Raman detuning it
obtains a more detailed structure -- the curvature at $k=0$ decreases, becomes
zero for $\Delta=\Gamma_{\hom},$ and turns negative for $\Delta>\Gamma_{\hom}$.

\begin{figure}
[ptb]
\begin{center}
\fbox{\includegraphics[
height=11.4246cm,
width=7.809cm
]%
{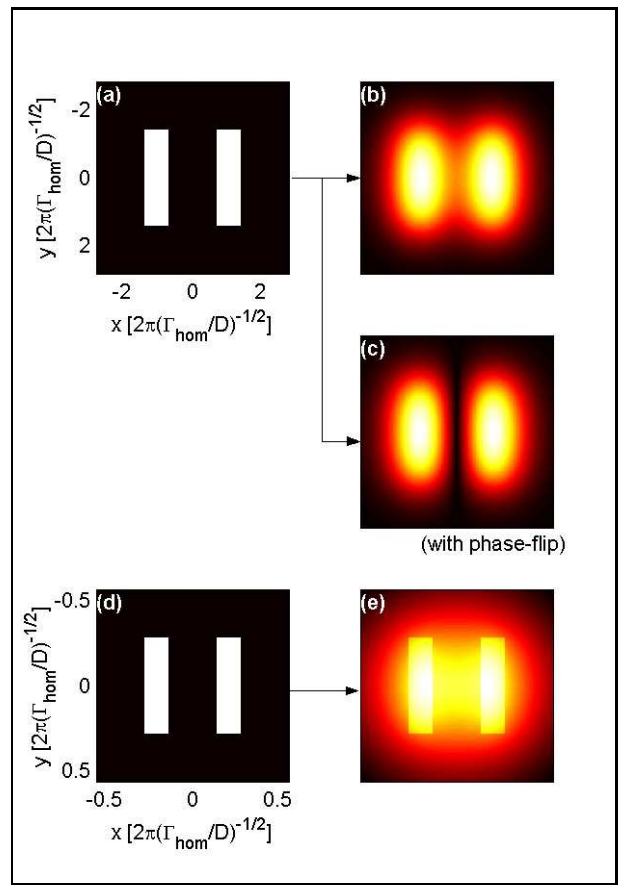}%
}\caption{Calculated effect of the spatial-frequency filter with a plane-wave
pump and a finite probe beam: The initial pattern, $|\tilde{\Omega}_{1}\left(
z_{1};x,y\right)  |$ (left), and the transmitted pattern, $|\tilde{\Omega}%
_{1}\left(  z_{2};x,y\right)  |$, after a certain propagation length (right).
The calculations were done using Eqs. (\ref{pk-1}), (\ref{ea-1}),
(\ref{chi-2}) and (\ref{Lkw}), and the diffraction ($k^{2}/2/q_{1}$) was
neglected for clarity. The parameters correspond to the black line in Fig.
\ref{fig_relk} ($K\left\vert \Omega_{2}\right\vert ^{2}=\Gamma_{12},$ $\delta
q=0,$ $\Delta=0$). Images (a)-(c) illustrate regular diffusion of real (b) and
complex (c) fields. To generate (c), the phase of the left line in the
incident field was flipped. Images (d) and (e) illustrate the effect for
smaller features, when larger $k_{\bot}$'s are pronounce.}%
\label{fig_images}%
\end{center}
\end{figure}

Figure \ref{fig_images} presents possible outcomes of $\tilde{\Omega}%
_{1}\left(  z_{2};x,y\right)  $, for several $\tilde{\Omega}_{1}\left(
z_{1};x,y\right)  $ (representing the absolute value of $\tilde{\Omega}_{1}$
as two-dimensional images). When the features in the incident image are large,
namely when $\tilde{\Omega}_{1}\left(  z_{1};\mathbf{k}_{\bot}\right)  $ is
confined within $k_{\bot}\ll k_{\text{typ}}$, the filter is approximately
\emph{quadratic in }$k_{\bot}$ (the central part of the solid-black line in
Fig. \ref{fig_relk}), which results in \emph{pure diffusion} in real space.
This is demonstrated in Figs. \ref{fig_images}(a) and \ref{fig_images}(b),
where the feature-size is of the order of $2\pi/k_{\text{typ}}$. In Fig.
\ref{fig_images}(c) we demonstrate the property of \emph{complex} diffusion --
the phase of the left line in $\tilde{\Omega}_{1}\left(  z_{1};x,y\right)  $
was shifted by $\pi,$ causing a \emph{destructive} \emph{interference} between
atoms that diffuse to the area between the lines, keeping it dark indefinitely
\cite{ShukerImaging2007}. When smaller features exist and $k_{\bot}$ extends
beyond $k_{\text{typ}}$ a more elaborate behavior occurs. Figure
\ref{fig_images}(d) is a small-scale version of \ref{fig_images}(a) and we
see, by comparing \ref{fig_images}(e) and \ref{fig_images}(b), that the
Lorentzian-shaped filter preserves the sharp edges in the smaller image. This
is due to the substantial deviation of the Lorentzian filter from a
quadratic-shaped filter (pure diffusion) for higher $k$'s.%

%TCIMACRO{\FRAME{ftbpFUX}{7.809cm}{11.4246cm}{0pt}{\Qcb{Calculated effect of
%the spatial-frequency filter with a plane-wave pump and a finite probe beam:
%The initial pattern, $|\tilde{\Omega}_{1}\left(  z_{1};x,y\right)  |$ (left),
%and the transmitted pattern, $|\tilde{\Omega}_{1}\left(  z_{2};x,y\right)  |$,
%after a certain propagation length (right). The calculations were done using
%Eqs. (\ref{pk-1}), (\ref{ea-1}), (\ref{chi-2}) and (\ref{Lkw}), and the
%diffraction ($k^{2}/2/q_{1}$) was neglected for clarity. The parameters
%correspond to the black line in Fig. \ref{fig_relk} ($K\left\vert \Omega
%_{2}\right\vert ^{2}=\Gamma_{12},$ $\delta q=0,$ $\Delta=0$). Images (a)-(c)
%illustrate regular diffusion of real (b) and complex (c) fields. To generate
%(c), the phase of the left line in the incident field was flipped. Images (d)
%and (e) illustrate the effect for smaller features, when larger $k_{\bot}$'s
%are pronounce.}}{\Qlb{fig_images}}{images.eps}%
%width 7.809cm;  height 11.4246cm;  depth 0pt;  original-width 10.3197cm;
%original-height 15.1391cm;  cropleft "0";  croptop "1";  cropright "1";
%cropbottom "0";  filename 'images.eps';file-properties "XNPEU";}}}%
%BeginExpansion
%EndExpansion

A direct measurement of the filter was carried out in Refs.
\cite{WeitzPRA2005,ShukerPRA2007}. In these experiments, the transmission of a
stationary $\left(  \omega=0\right)  $ on-resonance $\left(  \Delta=0\right)
$ probe beam was measured in two cases: an almost plane-wave probe,
propagating with a small angular deviation from the pump $\left(
\delta\mathbf{q}\neq0~;~\mathbf{k}=0\right)  ,$ and a divergent probe $\left(
\delta\mathbf{q}=0~;~\mathbf{k\neq0}\right)  .$ In both cases, the wave-number
was written as a function of the angular deviation ($\theta$), as $\left\vert
\mathbf{\delta q+k}\right\vert =q\theta,$ and the measured transmission agreed
with%
\begin{equation}
\Re eL\left(  \mathbf{k}\right)  =\frac{K|\Omega_{2}|^{2}}{\Gamma_{\hom
}+Dq^{2}\theta^{2}}.\label{ReL_theta}%
\end{equation}
We observe that, in the diverging beam experiment \cite{ShukerPRA2007}, the
outer parts of the beam are absorbed and consequently the beam radius
decreases. This result is counterintuitive from the viewpoint of diffusion,
since diffusion usually results in spreading, rather than contraction, of an
initial profile. Nevertheless, in the process of \emph{complex} diffusion, a
reduction in size can occur through destructive interference -- the diffusion
wipes out the field in regions where the phase pattern exhibits rapid changes.
The relative phase between a diverging Gaussian probe beam and a plane-wave
pump beam exhibits rapid variations in the transverse direction as the axial
distance increases, and thus diffusion results in a gradual elimination of the
outer parts of the beam.

\subsection{Diffusion of slow-light}

To conclude this section, we analyze the non-stationary behavior of slow-light
propagation. We consider the case of colinear pump and probe, $\mathbf{q}%
_{1}\parallel\mathbf{q}_{2}\parallel\mathbf{\hat{z}}$, and assume that the
changes in the probe's envelope along $z$ are much smaller than the changes in
the transverse plane, such that $\delta\mathbf{q}\cdot\mathbf{k\ll}k^{2}$ (or
alternatively, take $\delta\mathbf{q}=0$). Denoting $\mathbf{k=k}_{\perp
}+k_{z}\mathbf{\hat{z}}$, with $k_{z}\ll k_{\perp},$ Eq.(\ref{Lkw}) becomes
\begin{equation}
L=\frac{-K|\Omega_{2}|^{2}}{i\left(  \Delta+\omega\right)  -\Gamma_{\hom
}-D\delta q^{2}-Dk_{\perp}^{2}}.
\end{equation}
We further take the standard slow-light assumption, that the band-width of the
probe pulse is fully within the linear dispersion regime, i.e. that
$|\Delta+\omega|\ll|\Gamma_{\hom}+D\delta q^{2}|$. More importantly, we assume
that $Dk_{\perp}^{2}\ll\Gamma_{\hom}+D\delta q^{2}$, i.e. that the correction
to the EIT width resulting from the finite size of the probe is much smaller
than the EIT width of a plane-wave probe. The latter condition prevents the
dispersion of different spatial-frequency components of the envelope, and is
the essence of the diffusion approximation, allowing us to write $L$ as
quadratic in $k_{\perp}^{2}$,
\begin{equation}
L\approx\frac{K|\Omega_{2}|^{2}}{\Gamma_{\hom}+D\delta q^{2}}\left(
1+\frac{i\left(  \Delta+\omega\right)  -Dk_{\perp}^{2}}{\Gamma_{\hom}+D\delta
q^{2}}\right)  .\label{L-approx}%
\end{equation}

We return to the envelope equations of the probe, Eqs.(\ref{pk-1}%
)-(\ref{Ert-6}), and use the susceptibility of Eq.(\ref{chi-2}),%
\begin{equation}
\left[  k_{z}-\frac{\omega}{c}+\frac{k_{\perp}^{2}}{2q_{1}}-i\frac{g}{c}%
Kn_{0}\left(  1-L\right)  \right]  \tilde{\Omega}_{1}(\mathbf{k},\omega)=0.
\end{equation}
Substituting $L$ and defining the group-velocity $V_{g}$ as%
\begin{equation}
\frac{c}{V_{g}}=1+\frac{gn_{0}K^{2}|\Omega_{2}|^{2}}{\left(  \Gamma_{\hom
}+D\delta q^{2}\right)  ^{2}},
\end{equation}
we obtain%
\begin{align}
& \left[  ik_{z}-\frac{i\omega}{V_{g}}+\frac{ik_{\perp}^{2}}{2q_{1}}%
+\frac{gn_{0}}{c}K-\right. \\
& \left.  \left(  \frac{1}{V_{g}}-\frac{1}{c}\right)  \left(  \Gamma_{\hom
}+D\delta q^{2}-Dk_{\perp}^{2}+i\Delta\right)  \right]  \tilde{\Omega}%
_{1}(\mathbf{k},\omega)=0.\nonumber
\end{align}
Returning to the time and space coordinates and assuming $V_{g}\ll c$ for
brevity, we find%
\begin{equation}
\left[  V_{g}\frac{\partial}{\partial z}+\frac{\partial}{\partial t}-\left(
i\frac{V_{g}}{2q_{1}}+D\right)  \nabla_{\perp}^{2}+\Gamma_{0}-i\Delta\right]
\tilde{\Omega}_{1}\left(  \mathbf{r},t\right)  =0,
\end{equation}
where $\Gamma_{0}=$ $V_{g}gn_{0}K/c-\Gamma_{\hom}-D\delta q^{2}$ is the
on-resonance decay rate and $\nabla_{\perp}^{2}$ is the Laplacian
perpendicular to the $z-$axis. Introducing the \emph{traveling envelope} of
the probe beam, $\tilde{\Omega}_{1}^{\text{trav}}\left(  \mathbf{r},t\right)
$, as%
\begin{equation}
\tilde{\Omega}_{1}\left(  \mathbf{r},t\right)  =\tilde{\Omega}_{1}%
^{\text{trav}}\left(  \mathbf{r}-\mathbf{\hat{z}}V_{g}t,t\right)  e^{\left(
i\Delta-\Gamma_{0}\right)  t},
\end{equation}
we find that it undergoes a simple diffusion equation with a non-real
coefficient:%
\begin{equation}
\frac{\partial}{\partial t}\tilde{\Omega}_{1}^{\text{trav}}\left(
\mathbf{r},t\right)  =\left(  i\frac{V_{g}}{2q_{1}}+D\right)  \nabla_{\perp
}^{2}\tilde{\Omega}_{1}^{\text{trav}}\left(  \mathbf{r},t\right)
.\label{high_pb_solution}%
\end{equation}
According to Eq.(\ref{high_pb_solution}), a probe field with an arbitrary
complex envelope that satisfies the slow-light assumptions will undergo both
diffusion, as a result of the atomic thermal motion, and optical diffraction.
The diffraction depends on the actual distance traveled by the beam (due to
the factor $V_{g}/c$), while the diffusion depends on the time duration.

An interesting and important example is the propagation of a paraxial Gaussian
beam, such as a Hermite-Gaussian or a Laguerre-Gauss mode \cite{SiegmanBook}.
These modes have the well-known property of being self-similar under
diffraction, i.e. during the propagation through a diffractive medium their
transverse shape remains unchanged up to a length-scale factor. It can be
shown from Eq.(\ref{high_pb_solution}), by utilizing the "complex scaling
factor" representation \cite{SiegmanBook}, that Gaussian modes are also
self-similar under diffusion. For example, the lowest order mode has a
Gaussian intensity profile, which is known to maintain a Gaussian shape when
diffusing. It can further be shown from Eq.(\ref{high_pb_solution}) that the
Gaussian modes experience a diffusion-induced decay throughout the
propagation, as discussed at the end of \S \S \ref{ss_storage}.

\section{Finite Pump and Probe Beams\label{ch_ramsey}}

When both the probe and the pump beams are finite, atoms can leave the light
beams, evolve "in the dark" and diffuse back inside. It was recently
demonstrated that such a process may result in an EIT line much narrower than
expected from time-of-flight (TOF) broadening and power broadening -- a
phenomenon denoted as Ramsey narrowing \cite{WalsworthPRL2006}. The line
shapes resulting from TOF broadening and Ramsey narrowing can be described by
following the possible atomic paths (trajectories), calculating temporal
probability functions for the atoms' location, and averaging over them (cf.
\cite{WalsworthPRL2006}). If the time it takes to achieve steady-state with
the driving field (pumping rate) is comparable to the TOF, it is commonly
claimed that calculations cannot be done with standard steady-state approaches
and time-dependent solutions have to be used \cite{Gawlik1986}. Nevertheless,
here we calculate these effects using the steady-state solution of the
diffusion equation in the presence of the fields. This is an exact and easier
approach that allows more elaborate beam geometries to be considered. Note
that the Ramsey-narrowing experiments are usually done with equal pump and
probe intensities, while our model is for the weak probe regime. However, we
expect the main attributes of the spectrum to be essentially the same for both cases.

We consider finite probe and pump beams and restrict the discussion to a
colinear EIT, $\mathbf{q}_{1}-\mathbf{q}_{2}=\delta q\mathbf{\hat{z}.}$ We
assume that the fields are stationary and overlap in their cross sections with
a neglected variation along the $z-$direction,%
\begin{equation}
\tilde{\Omega}_{1}\left(  \mathbf{r},t\right)  =\Omega_{1}w\left(
\mathbf{r}_{\bot}\right)  ~~;~~\tilde{\Omega}_{2}\left(  \mathbf{r}\right)
=\Omega_{2}w\left(  \mathbf{r}_{\perp}\right)  ,
\end{equation}
with $w\left(  \mathbf{r}_{\perp}\right)  $\ the transverse profile of the
fields. In the diffusion regime we use Eqs.(\ref{ac-6}) and (\ref{ac-2}),
which can now be written as%
\begin{align}
& \left[  \Gamma-i\Delta+K|\Omega_{2}|^{2}|w\left(  \mathbf{r}_{\perp}\right)
|^{2}-D\nabla_{\perp}^{2}\right]  \tilde{R}_{21}\left(  \mathbf{r}_{\perp
}\right)  =\nonumber\\
& -n_{0}K\Omega_{2}^{\ast}\Omega_{1}|w\left(  \mathbf{r}_{\perp}\right)
|^{2},\label{b-2}%
\end{align}
and%
\begin{equation}
\tilde{R}_{31}(\mathbf{r}_{\perp})=iK\left[  \Omega_{2}\tilde{R}%
_{21}(\mathbf{r}_{\perp})+\Omega_{1}n_{0}\right]  w\left(  \mathbf{r}_{\perp
}\right)  ,\label{b-1}%
\end{equation}
where we denoted $\Gamma=\Gamma_{21}+D\delta q^{2}$ to be the
non-power-broadened width. In what follows, we solve for $\tilde{R}%
_{21}\left(  \mathbf{r}_{\perp}\right)  $ and $\tilde{R}_{31}\left(
\mathbf{r}_{\perp}\right)  $ in a specific example and calculate the resulting
absorption spectrum.

\subsection*{Example: a Stepwise Beam}

We consider a probe and a pump beams with uniform intensity and phase within a
\emph{sheet} of thickness $2a$ in the $x-$direction (one-dimensional stepwise
beams):%
\begin{equation}
w\left(  x,y\right)  =\left\{
\begin{array}
[c]{c}%
1~~\ \text{for }\left\vert x\right\vert \leq a\\
0~~\ \text{for }\left\vert x\right\vert >a
\end{array}
\right.  .\label{b-3}%
\end{equation}
Eq.(\ref{b-2}) can then be written as%
\begin{align}
D\left(  k_{1}^{2}-\frac{\partial^{2}}{\partial x^{2}}\right)  \tilde{R}%
_{21}\left(  \left\vert x\right\vert \leq a\right)   & =-n_{0}K\Omega
_{2}^{\ast}\Omega_{1},\nonumber\\
D\left(  k_{2}^{2}-\frac{\partial^{2}}{\partial x^{2}}\right)  \tilde{R}%
_{21}\left(  \left\vert x\right\vert >a\right)   & =0,
\end{align}
where%
\begin{align}
k_{1}  & =\sqrt{(\Gamma+K|\Omega_{2}|^{2}-i\Delta)/D},\nonumber\\
k_{2}  & =\sqrt{(\Gamma-i\Delta)/D},\label{k1_k2}%
\end{align}
and $\Re e\left\{  k_{i}\right\}  >0$. For $|x|\leq a$ we expect a solution
symmetric in $x$ and for $\left\vert x\right\vert >a$ we expect a solution
decaying for $\left\vert x\right\vert \rightarrow\infty$. We thus find
\begin{align}
\tilde{R}_{21}(\left\vert x\right\vert  & \leq a)=A\cosh\left(  k_{1}x\right)
-\frac{Kn_{0}}{k_{1}^{2}D}\Omega_{2}^{\ast}\Omega_{1},\nonumber\label{b-8}\\
\tilde{R}_{21}(|x|  & >a)=B\exp\left[  -k_{2}\left(  \left\vert x\right\vert
-a\right)  \right]  ,
\end{align}
and the coefficients $A$ and $B$ are obtained from the continuity conditions
of $\tilde{R}_{21}$ and $\frac{\partial}{\partial x}\tilde{R}_{21}$ at
$\left\vert x\right\vert =a$:
\begin{align}
A  & =\frac{Kn_{0}}{k_{1}^{2}D}\frac{\Omega_{2}^{\ast}\Omega_{1}}{\cosh
(k_{1}a)+\left(  k_{1}/k_{2}\right)  \sinh(k_{1}a)},\nonumber\\
B  & =A\cosh\left(  k_{1}a\right)  -\frac{Kn_{0}}{k_{1}^{2}D}\Omega_{2}^{\ast
}\Omega_{1}.
\end{align}

\begin{figure}
[ptb]
\begin{center}
\includegraphics[
height=7.44cm,
width=8.1121cm
]%
{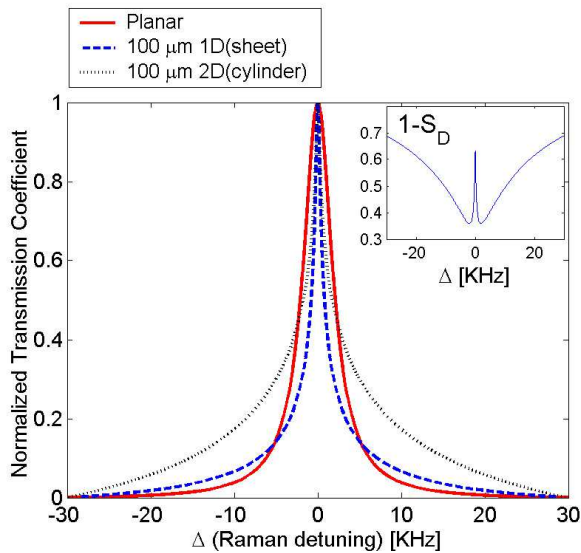}%
\caption{Normalized transmission for a plane-wave and a finite-sized beam (1D
and 2D), demonstrating TOF broadening and Ramsey-narrowing. The inset depicts
the correction to the spectrum, resulting from the finiteness of beams, $\Re
e\left[  1-S_{D}\left(  \Delta\right)  \right]  $, for the 1D case
[Eq.(\ref{sd1d})]. The parameters are: $\Gamma=100$ Hz, $K|\Omega_{2}|^{2}=2$
KHz, $D=10$ cm$^{2}/$sec and $a=100$ $\mu$m. The choice of $K|\Omega_{2}%
|^{2}\gg\Gamma$ makes the narrowing effect more obvious.}%
\label{fig_Ramsey}%
\end{center}
\end{figure}

For $\tilde{R}_{31}(x)$ we find from Eq.(\ref{b-1}) that $\tilde{R}%
_{31}(\left\vert x\right\vert >a)=0$ and
\begin{align}
\tilde{R}_{31}(\left\vert x\right\vert  & <a)=i\Omega_{1}n_{0}K\left(
1-\frac{K|\Omega_{2}|^{2}}{k_{1}^{2}D}\right) \nonumber\\
& ~~~~~~~+iKA\Omega_{2}\cosh\left(  k_{1}x\right)  .
\end{align}
The energy absorption at frequency $\omega_{1}$ is%
\begin{equation}
P(\Delta)=2\hbar\omega_{1}\frac{1}{2a}\int_{-a}^{a}dx\Im m\left\{  \Omega
_{1}^{\ast}\tilde{R}_{31}(x)\right\} \label{pl-1}%
\end{equation}
and we find
\begin{equation}
P(\Delta)=P_{0}\Re e\left\{  K-\frac{K^{2}|\Omega_{2}|^{2}}{\Gamma
+K|\Omega_{2}|^{2}-i\Delta}\left[  1-S_{D}\left(  \Delta\right)  \right]
\right\}  ,\label{P_delta}%
\end{equation}
where $P_{0}=2\hbar\omega_{1}n_{0}\left\vert \Omega_{1}\right\vert ^{2}$ and
\begin{equation}
S_{D}\left(  \Delta\right)  =\frac{\tanh(k_{1}a)}{k_{1}a}\frac{1}{1+\left(
k_{1}/k_{2}\right)  \tanh(k_{1}a)}\label{sd1d}%
\end{equation}
is the correction resulting from the finite size of the beam. Figure
\ref{fig_Ramsey} depicts $P(\Delta)$ and $\Re eS_{D}\left(  \Delta\right)  $
for $a=100\mu$m and $a\rightarrow\infty$ (plane-wave) with typical parameters.
The outer part of the finite-beam spectrum (dashed-blue) is broadened due to
the TOF effect. Ramsey-narrowing is apparent in the central part as a
cusp-like curve. The cusp is narrower than the power-broadened Lorentzian, but
it is nevertheless limited by the width $\Gamma=\Gamma_{21}+D\delta q^{2}$.%
%TCIMACRO{\FRAME{ftbpFU}{8.1121cm}{7.44cm}{0pt}{\Qcb{Normalized transmission
%for a plane-wave and a finite-sized beam (1D and 2D), demonstrating TOF
%broadening and Ramsey-narrowing. The inset depicts the correction to the
%spectrum, resulting from the finiteness of beams, $\Re e\left[  1-S_{D}\left(
%\Delta\right)  \right]  $, for the 1D case [Eq.(\ref{sd1d})]. The parameters
%are: $\Gamma=100$ Hz, $K|\Omega_{2}|^{2}=2$ KHz, $D=10$ cm$^{2}/$sec and
%$a=100$ $\mu$m. The choice of $K|\Omega_{2}|^{2}\gg\Gamma$ makes the narrowing
%effect more obvious.}}{\Qlb{fig_Ramsey}}{ramsey.eps}%
%width 8.1121cm;  height 7.44cm;  depth 0pt;  original-width 14.0057cm;
%original-height 12.8349cm;  cropleft "0";  croptop "1";  cropright "1";
%cropbottom "0";  filename 'Ramsey.eps';file-properties "XNPEU";}}}%
%BeginExpansion
%EndExpansion

A similar calculation can be done for the two-dimensional analogue of the
stepwise sheet: a stepwise cylindrical profile with $w\left(  r\leq a\right)
=1$ and $w\left(  r>a\right)  =0$, where $r^{2}=x^{2}+y^{2}$. Solving Eqs.
(\ref{b-2}) and (\ref{b-1}) in cylindrical symmetry, one finds that the energy
absorption spectrum, $P(\Delta)$, has the exact same form of Eq.(\ref{P_delta}%
), with the correction term being%
\begin{equation}
S_{D}\left(  \Delta\right)  =\frac{2}{k_{1}a}\left[  \frac{I_{0}\left(
k_{1}a\right)  }{I_{1}\left(  k_{1}a\right)  }+\frac{k_{1}}{k_{2}}\frac
{K_{0}\left(  k_{2}a\right)  }{K_{1}\left(  k_{2}a\right)  }\right]  ^{-1}.
\end{equation}
Here, $k_{1,2}$ are as defined in Eqs.(\ref{k1_k2}) and $I_{0}\left(
x\right)  $ and $K_{0}\left(  x\right)  $ are modified Bessel functions. A
comparison between the 1D and the 2D spectra is given in Fig. \ref{fig_Ramsey}%
. In the 2D case the TOF effect is substantial while the Ramsey-narrowing is
reduced. The latter can be attributed to the fact that on average less atoms
return to the beam in the 2D geometry.

\section{Conclusions}

We have presented a model for EIT that incorporates thermal atomic motion by
introducing the density-matrix distribution in space and velocity along with a
Boltzmann relaxation term. The model describes a range of motional phenomena,
such as Dicke narrowing, Ramsey narrowing and diffusion during storage of
light, which have been analyzed in the past in different independent studies.
In the absence of electromagnetic fields, the model reduces to pure diffusion
of the ground-state coherence and population, in agreement with recent storage
of light experiments. In the presence of a pump and a weak probe fields, the
solution for the spatial distribution of the atomic coherence and the fields
can be obtained.

For the case of a plane-wave pump, the \emph{temporal and spatial frequency
components} of the incoming probe beam form the natural basis for the problem.
The general solution, in terms of the complex susceptibility of the medium,
$\chi_{31}(\mathbf{k},\omega)$, exhibits the Doppler-Dicke transition for both
the one-photon and the two-photon absorption spectra. For the realistic
regime, in the presence of a buffer-gas, when the one-photon line is
Doppler-broadened and the two-photon line is Dicke-narrowed, an explicit
expression for the EIT transmission, $L(\mathbf{k},\omega)$, is derived.
$L(\mathbf{k},\omega)$ yields the EIT absorption spectrum, for any given
$\mathbf{k}$, including the power-broadening effect and the Dicke width.
Moreover, for any given $\omega$, $L(\mathbf{k},\omega)$ serves as a
spatial-frequency filter, generally diminishing high $k$ values. We explain
this diminution by the diffusion of atoms across the pump-probe interference
pattern, of wave-vector $\mathbf{k}+\mathbf{q}_{1}-\mathbf{q}_{2}$. On
Raman-resonance, and when the spatial features of the incoming probe beam are
large, the probe's envelope undergoes a diffusion-like dynamics. For smaller
features or for non-zero Raman-detuning, a more elaborate behavior takes
place. We note that a unique result is obtained for probe beams with a single
value of $|\mathbf{k}_{\bot}|$, usually referred to as non-diffracting, e.g.
Bessel beams \cite{DurninPRL1987}. These will not be distorted by
$L(\mathbf{k},\omega)$ and hence will not spread due to neither diffusion nor diffraction.

Since the \emph{complex} amplitude of the probe's envelope diffuses,
interference phenomena occur. For example, destructive interference between
adjacent features that are opposite in phase, maintains the dark area between
them. This also explains why adjacent rings in the Bessel beam remain
separated indefinitely. Furthermore, when the optical diffraction is taken
into account, the effective diffusion coefficient becomes a complex number
($D+iV_{g}/2/q_{1}$), with the group-velocity determining the ratio between
the real (pure-diffusion) and the imaginary (diffraction) parts. Such complex
diffusion can possibly be useful for all-optical image processing, such as
image enhancement, denoising and edge-detection \cite{ZeeviIEEE2004}.

Ramsey narrowing occurs when the pump's cross-section is finite, and atoms
that re-enter the beam from outside, less affected by power-broadening,
contribute to the spectrum. Our model gives simple and analytic results for
the Ramsey-narrowed spectrum by solving a diffusion equation with spatially
dependent decay coefficients and sources. It is somewhat surprising that our
steady-state approach is able to capture this effect, for any pumping and
transit rates, so that one is not required to average over atomic
trajectories. Utilizing our model, the spectrum for any detailed geometry can
readily be obtained.

The theory presented here may contribute to the analysis of decoherence in
collective light memories, for which it was shown that the decay rate is
proportional to the single-atom diffusion rate \cite{MewesFleischhauer2005}.
The model can potentially be extended to include the main ground-state
decoherence mechanisms in vapor: non-coherence-preserving collisions, namely
spin-exchange collisions, and wall collisions. Such extensions may aid in
developing vapor EIT schemes with narrower lines.

\begin{acknowledgments}
This work was partially supported by DDRND\ and the fund for encouragement of
research in the Technion.
\end{acknowledgments}

\bibliographystyle{apsrev}
\bibliography{proposal_additional,references}

\end{document}